\documentclass[12pt,eqsecnum,nofootinbib,floats,article,aps,prd,floatfix,titlepage,superscriptaddress]{revtex4} 
\usepackage{epsfig}
\usepackage{graphicx}
\usepackage{bm}
\usepackage{amsmath}
\usepackage{amssymb}
\usepackage[multiple]{footmisc}
\usepackage{setspace}

\usepackage{epstopdf}
\begin{document}

\title{Kink Collisions in Curved Field Space}

\author{Pontus Ahlqvist}
\email{pontus@phys.columbia.edu}
\affiliation{Physics Department, Columbia University, New York, New York 10027, USA}
\author{Kate Eckerle}
\email{ke2176@columbia.edu}
\affiliation{Department of Physics and Department of Applied Mathematics, Columbia University, New York, New York 10027, USA}
\author{Brian Greene}
\email{greene@physics.columbia.edu}
\affiliation{Department of Physics and Department of Mathematics, Columbia University, New York, New York 10027, USA}

\begin{abstract}

We study bubble universe collisions in the ultrarelativistic limit with the new feature of allowing for nontrivial curvature in field space. We establish a simple geometrical interpretation of such collisions in terms of a double family of field profiles whose tangent vector fields stand in mutual parallel transport. This provides a generalization of the well-known flat field space limit of the free passage approximation. We investigate the limits of this approximation and illustrate our analytical results with a numerical simulations.

\end{abstract}

\maketitle

\section{Introduction}

Some time ago, a series of seminal papers \cite{Coleman, ColemanCallan, ColemanDeLuccia}, showed that the universe can undergo quantum tunneling from one local minima of a potential to another. Much more recently, a classical mechanism for vacuum transitions was introduced by \cite{Easther; Giblin, Giblin} involving coherent collisions between bubble universes. While these processes, both quantum and classical, are of significant theoretical interest in and of themselves, their relevance may be far greater due to the fact that models of inflationary cosmology typically yield numerous expanding bubble universes whose centers of nucleation are sufficiently close to allow for such collisions. Moreover, string theory offers the possibility of an enormous landscape of local minima, making it important to determine if tunneling events and bubble collisions are essential cosmological processes. These considerations have inspired numerous authors to study transitions between collections of metastable vacua, with results of possible importance to  key outstanding issues, such as the cosmological constant problem and the search for experimental signatures of a multiverse  \cite{Susskind, BoussoPolch, persistence-memory, Aguirre-Johnson, Johnson, BrianAli, Shiu, Tye, BrownSarangi, Brown}.

In this paper, we focus on aspects of bubble universe collisions. An important observation in this regard was reported in \cite{Easther; Giblin, Giblin} which found that at sufficiently high relative velocity the physics vastly simplifies. Namely, at at high impact velocity, colliding bubble walls are so Lorentz contracted that the time it takes them to pass through one another is less than the time for interactions between the bubbles to contribute significantly. Thus, the field configurations in such situations simply superpose--the ``free passage" approximation. As the relative velocity of colliding bubbles is proportional to the initial separation between their centers of nucleation, the free passage approximation becomes ever more accurate for ever larger separations.

When free passage holds, there is thus a finite window of time during which the field's value in the widening spatial region through which both walls have passed in opposite directions -- the collision region -- stays nearly homogeneous. The field's value in this region is given by the ambient background (the ``parent field value") plus the sum of the field value changes across each bubble wall, i.e. the sum of the bubble field values minus the parent value. The general expectation is that post free passage, the field will be driven by the slope of the potential at this shifted field value, causing it to settle into the nearest local minimum. Thus if the free-passage-kicked field value is in the basin of attraction of a different local minimum (neither bubble, nor parent) then the collision will have spawned a new bubble universe in which the field has acquired the new value. To be sure, in \cite{us} we pointed out some subtleties in this picture (in which the strength of the forcing function at the free passage induced field value can be sufficient to pull the field out of the new basin of attraction, causing it to finally settle at the original field value and thus thwarting the creation of a new bubble universe), but we anticipate that for many situations, these subtleties will not arise.

To date, studies of bubble collisions have generally considered theories with a single real scalar field described by canonical kinetic terms, and for the most part ignoring gravity \cite{Easther; Giblin, Giblin, Aguirre-Johnson, Johnson} \footnote{A notable exception is the paper \cite{Johnson}}. In many common theoretical scenarios, however, bubble collisions occur in theories whose scalar fields parameterize a many-dimensional, curved manifold.  For instance, Calabi-Yau compactifications of string theory involve scalar fields which are local coordinates on moduli spaces that are generally complex K{\"a}hler manifolds with nontrivial curvature--so-called ``moduli fields". The string landscape is due, in part, to the various local minima of flux potentials that govern the dynamics of these moduli.

A natural issue, then, is the impact of a nontrivial metric on bubble universe collisions, which is the issue we take up in this paper.  We address key subtleties in bubble collisions, even at high relative velocity, that arise from the inherent nonlinearity of nontrivial curvature, and find a satisfying geometrical interpretation of our result. Specifically, in Section II we generalize the notion of the free passage approximation from flat to curved field spaces. We derive a geometrical interpretation of the result in terms of the parallel transport of integral curves on moduli space. In Section III we argue that there always exists a regime in which our generalized free passage approximation applies, and in Section IV we numerically study some bubble collision examples (in the setting of 1+1D) and compare the results to those from our analytic expressions in Section II. Finally, we end with some conclusions and suggestions for further work.

\section{Generalization of Free Passage}

We take as our starting point the action
\begin{equation}
S=\int_{}^{}d^2 x \left(\frac{1}{2}g_{ij}\partial_\mu\phi^i\partial^\mu\phi^j-V(\phi)\right)
\label{Action}
\end{equation}
where $g_{ij}$ is the generally curved metric on the field space. We assume the potential $V(\phi)$ has three (or more) degenerate minima at the field space locations $A$, $B$, and $C$ (the minimum necessary to study collisions as a source of vacuum transitions). The Euler-Lagrange equation takes the form,
\begin{equation}
\Box\phi^i +\Gamma^i_{j\thickspace k}\partial^\mu\phi^j\partial_\mu\phi^k =- \frac{\partial V}{\partial \phi_i}\label{curved-EOM}
\end{equation}
The $A$ vacuum will play the role of the parent vacuum, and $B$ and $C$ those of the two bubble vacua we seek to collide. Static solutions to (\ref{curved-EOM}) that interpolate between distinct yet degenerate local minima of the potential are solitons. We define $f^i(x)$, and $h^i(x)$ as the components of those solitons centered at $x=0$ with the following asymptotics,
\begin{align}
\lim_{x\to -\infty}f^i(x)&=B^i\\
\lim_{x\to +\infty}f^i(x)&=A^i\\
\lim_{x\to -\infty}h^i(x)&=C^i\\
\lim_{x\to +\infty}h^i(x)&=A^i
\label{solitonBCs}
\end{align}
Our intent is to work out the formalism to describe the collisions between these solitons, taking into account the curved moduli space metric. Of particular interest is the limiting behavior that emerges at ultrarelativistic impact velocity.

The collision of two initially widely separated solitons, say, right-moving $f^i$, and left-moving $h^i$ (that interpolate between the parent vacuum $A$  and the other local minima $B$ and $C$,  respectively)  is described by an observer in the center of the rest frame of the collision by the following initial value problem:
\begin{align}
&\Box\phi^i +\Gamma^i_{j\thickspace k}\partial^\mu\phi^j\partial_\mu\phi^k =- \frac{\partial V}{\partial \phi_i} \label{collisionIVP}\\
\phi^i(-T,x)&=f^i(\gamma(x-u(-T)))+h^i(-\gamma(x+u(-T)))-A^i\label{collisionIC1}\\
\partial_t \phi^i(-T,x)&=-u\gamma\left({f^i}'(\gamma(x-u(-T)))+{h^i}'(-\gamma(x+u(-T)))\right)\label{collisionIC2}
\end{align}
where we've shifted the time coordinate so that the observer's clock is zero when the trajectories of the centers of the colliding solitons (given by $x_{R,0}=ut$ for right-moving $f^i$, and $x_{L,0}=-ut$ for left-moving $h^i$) coincide.

In order to be a legitimate representation of a soliton collision, the solitons must be widely separated at the initial time, $-T$. Thus, valid values of $T$ are those for which $uT$ is much much greater than the width of all components of both Lorentz contracted solitons (so the observer measures field value $\phi^i=A^i$ to an exceedingly good approximation initially, since outside this width the solitons approach their asymptotic values as decaying exponentials). To make this precise we'll define a positive constant $w$ such that all components of the solitons we wish to collide, $f^i(x)$ and $h^i(x)$, differ from the relevant vacuum value by an insignificant amount outside of $x\in[-w/2,w/2]$. That is,
\begin{align}
&\frac{|B^i-f^i(-w/2)|}{|A^i-B^i|}\ll 1\\
&\frac{|A^i-f^i(w/2)|}{|A^i-B^i|}\ll 1\\
&\frac{|C^i-h^i(-w/2)|}{|A^i-C^i|}\ll 1\\
&\frac{|A^i-f^i(w/2)|}{|A^i-C^i|}\ll 1.
\end{align}

So, the initial time $-T$ is any time such that $uT> w/2\gamma$. We'll view the collision as commencing at time $t_{start}=-w/2\gamma$, and lasting until $t_{end}=+w/2\gamma$. The ``usefulness" then of an approximation to the actual solution to the collision initial value problem (defined by \ref{collisionIVP}, \ref{collisionIC1}, and \ref{collisionIC2}), at given impact velocity $u$, is proportional to the fraction of the collision for which the approximation remains valid. Labeling the time until which an approximation accurately captures the dynamics by $t_{approx}$, the approximation's usefulness is gleaned from $t_{approx}/t_{end}$. The greater this is the more useful the approximation is, and it will be deemed ``fully realized" if $t_{approx}\ge t_{end}$.

The task of understanding the dynamics of collisions in the ultrarelativistic limit amounts to finding a one parameter family of approximations-- ``free passage field configurations" we'll denote by $\{\phi^i_{FP}(t,x;u)\}_{u\in(0,1)}$-- that are ever more useful approximations to the collision initial value problem's true solution as $u$ is taken to $1$. After constructing  $\{\phi^i_{FP}(t,x;u)\}_{u\in(0,1)}$ we conclude this section with a proof that there always exists an impact velocity close enough to $1$ to ensure that the free passage configuration is fully realized. 

\begin{figure}
\includegraphics[width=.5 \textwidth]{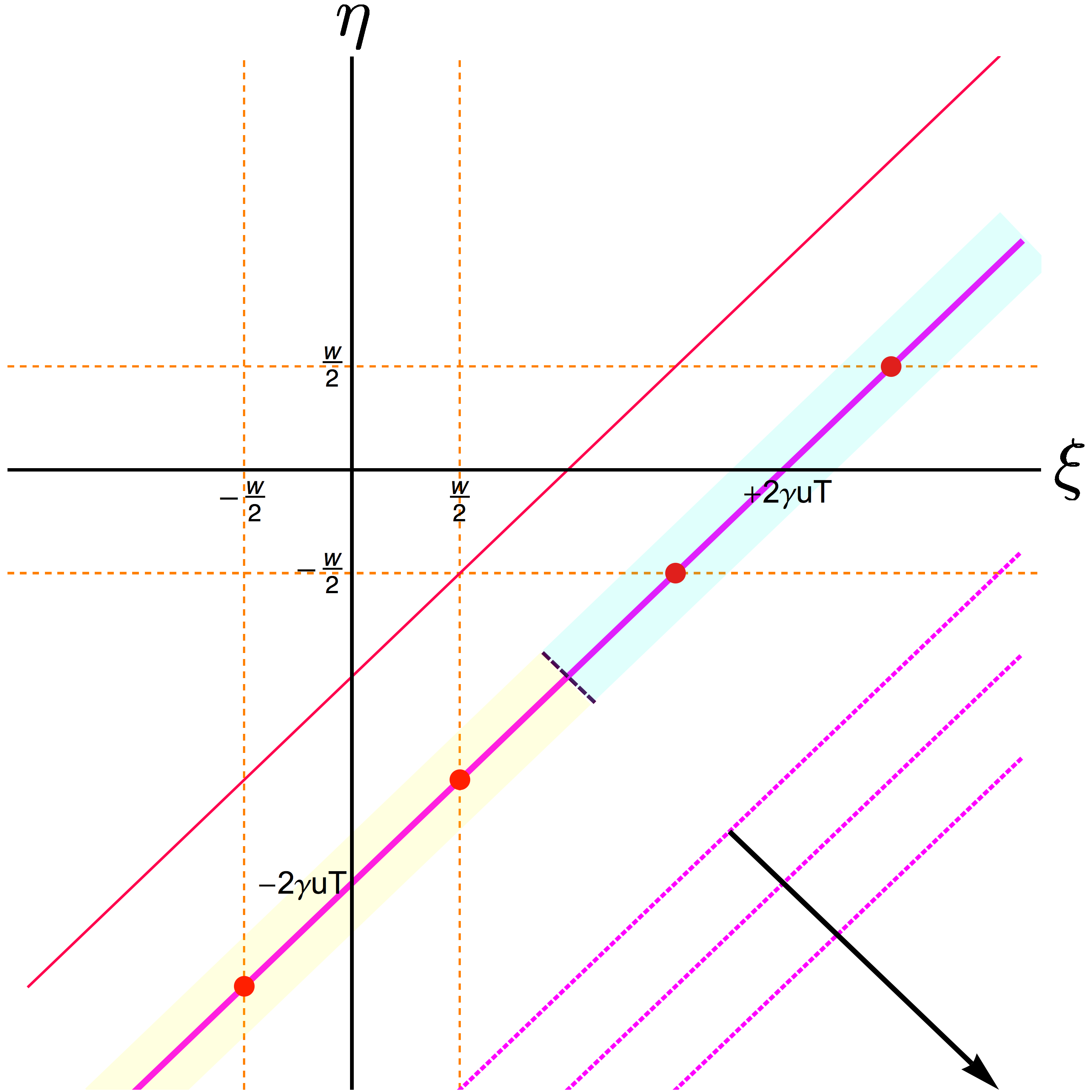}
\caption{The initial conditions for the collision of solitons boosted to impact velocity $u$ are given by \ref{initialdata1} and \ref{initialdata2} when a transformation from ($t$, $x$) to the Lorentz variables ($\sigma=\gamma(x-ut)$, $\omega=\gamma(x+ut)$)  is performed. This boundary where the initial data is given, the surface of constant time $t=-T$, is a line in the $\sigma$-$\omega$ plane with slope $1$ and $\omega$-intercept $-2\gamma u T$. We denote this boundary, for a given Lorentz factor and choice for $T$, by $\partial\Omega_\gamma$ (note that for the collision at a given $u$ the valid $T$ values are those such that the boundary lies below the line $\omega=\sigma-w$, indicated by the thin red line). The center of the $f$ soliton in the shifted superposition in the boundary conditions occurs at the $\omega$-intercept, and that of the $h$ soliton in the superposition occurs at the $\omega$-intercept. We've named the halves of $\partial\Omega_\gamma$ as $\partial\Omega_f$ and $\partial\Omega_h$ according to which soliton's center lies on it. These are indicated by the yellow and blue rays, respectively. In fact, the $f$ soliton in the superposition is almost entirely contained within only the two red points on the yellow half-- the points on $\partial\Omega_\gamma$ with $\sigma\in[-w/2,w/2]$. That is, to the left of this interval (in the third quadrant) $f$ evaluates to very nearly $B$, and evaluates to very nearly $A$ to the left. Similarly the $h$ soliton is almost entirely contained within the two red points on the blue half. As $\gamma$ is increased the boundary where the field and its derivatives are specified moves along the diagonal with negative slope toward the fourth quadrant, indicated by the black arrow. Consequently, the intervals where the $f$ and $h$ solitons in the superposition are approximately supported move further and further away from each other. Consequently, the boundary data effectively splits into two independent pieces each involving a \emph{single} Lorentz variable. On the yellow half boundary the field in the initial data evaluates ever more closely to $f(\sigma)$, and on the blue to $h(-\omega)$. In the limit $\gamma\rightarrow 1$ the $\omega$ values on the yellow half go to $-\infty$, and the $\sigma$ values on the blue half go to $+\infty$.}
\label{boundary-split}
\end{figure}

We first perform a change of variables from $t$ and $x$ to the natural dynamical variables of the problem, namely the spatial coordinates of the boosted observers riding on the soliton walls, $\sigma\equiv\gamma(x-ut)$, and $\omega\equiv\gamma(x+ut)$, which we'll refer to as the Lorentz variables. This choice of variables enables us to isolate the effect of one soliton, say, left-moving $h$, on the field at a fixed location on the other soliton, in this case $f$. By holding $\sigma$ constant and letting $\omega$ vary from minus infinity to infinity one focuses on a fixed location on the right-moving soliton and follows how the field evolves under the influence of the collision with the left-moving soliton. Similarly, the impact of right-moving $f$ on $h$ can be ascertained by holding $\omega$ constant and varying $\sigma$. Expressing $\phi$ in terms of these, the equation of motion takes the form,
\begin{align}
-4(1-\epsilon)\gamma^2\left[\frac{\partial^2 {\phi}^i}{\partial\sigma\partial\omega}+\Gamma^i_{j\thickspace k}\frac{\partial{\phi}^j}{\partial\sigma}\frac{\partial \phi^k}{\partial\omega} \right]-2\gamma^2\epsilon\left[\frac{\partial^2\phi^i}{\partial\sigma^2}+\frac{\partial^2\phi^i}{\partial\omega^2}+\Gamma^i_{j\thickspace k} \left(\frac{\partial{\phi}^j}{\partial\sigma}\frac{\partial{\phi}^k}{\partial\sigma}+ \frac{\partial{\phi}^j}{\partial\omega}\frac{\partial{\phi}^k}{\partial\omega}\right)\right]= -\frac{\partial V}{\partial {\phi}_i}\label{LorentzVarsEOM}
\end{align}

where we've expanded in $\epsilon=1-u$, since we are interested in the limiting dynamics that emerge when $u\rightarrow 1$. Rearranging and using $1/\gamma^2=2\epsilon$ we have,
\begin{align}
\frac{\partial^2 {\phi}^i}{\partial\sigma\partial\omega}+\Gamma^i_{j\thickspace k}\frac{\partial{\phi}^j}{\partial\sigma}\frac{\partial \phi^k}{\partial\omega}=\frac{\epsilon}{2}\frac{\partial V}{\partial {\phi}_i}-\frac{\epsilon}{2}\left[\Gamma^i_{j\thickspace k} \left(\frac{\partial{\phi}^j}{\partial\sigma}\frac{\partial{\phi}^k}{\partial\sigma}+ \frac{\partial{\phi}^j}{\partial\omega}\frac{\partial{\phi}^k}{\partial\omega}\right)+\frac{\partial^2\phi^i}{\partial\sigma^2}+\frac{\partial^2\phi^i}{\partial\omega^2}\right].\label{EPSparallelPDE}
\end{align}
The initial conditions take the form,
\begin{align}
\phi^i(\sigma,\omega)\bigg|_{\partial\Omega_\gamma}=&\left(f^i(\sigma)+h^i(-\omega)-A^i\right)\bigg|_{\partial\Omega_\gamma}\\
\gamma u\left(\frac{\partial}{\partial\omega}-\frac{\partial}{\partial\sigma}\right)\phi^i\bigg|_{\partial\Omega_\gamma}=&-\gamma u\left[{f^i}'(\sigma)+{h^i}'(-\omega)\right]\bigg|_{\partial\Omega_\gamma}
\end{align}
or, 
\begin{align}
\phi^i(\sigma,\omega)\bigg|_{\partial\Omega_\gamma}=&\left(f^i(\sigma)+h^i(-\omega)-A^i\right)\bigg|_{\partial\Omega_\gamma}\label{initialdata1}\\
\left(\frac{\partial}{\partial\omega}-\frac{\partial}{\partial\sigma}\right)\phi^i\bigg|_{\partial\Omega_\gamma}=&-\left[{f^i}'(\sigma)+{h^i}'(-\omega)\right]\bigg|_{\partial\Omega_\gamma}\label{initialdata2}
\end{align}
where $\partial\Omega_\gamma$ is the surface in the $\sigma$-$\omega$ plane of constant time $t=-T$. This boundary is simply the line, $\omega=\sigma-2\gamma uT$, which note lies only in the first, third, and fourth quadrants. Its $\omega$-intercept, $-2\gamma uT$, is less than $-w$ for any valid choice of $T$. We bisect $\partial\Omega_\gamma$ at the point $(\gamma uT,-\gamma uT)$ and name the half that lies in the third and lower fourth quadrants as $\partial\Omega_f$, and the half that lies in the first and upper fourth quadrants as $\partial\Omega_h$. These are indicated in Figure \ref{boundary-split} by the highlighted yellow, and blue rays, respectively. 

Since all points on $\partial\Omega_f$ have $\sigma<\gamma u T<w/2$, they satisfy,

\begin{align*}
\rightarrow \omega\bigg|_{\partial\Omega_f}=(\sigma-2\gamma u T)\bigg|_{\partial\Omega_f} \leq-\gamma u T\leq -w/2
\end{align*}
Similarly, all points on $\partial\Omega_h$ have $\omega>-\gamma u T>-w/2$, so their corresponding $\sigma$ coordinate satisfies,
\begin{align*}
\sigma\bigg|_{\partial\Omega_h}&\geq2\gamma u T\geq w/2
\end{align*}
The boundary conditions can then be rewritten as, 
\begin{align*}
\partial\Omega_f:&\thickspace \sigma<w/2\rightarrow \sigma-2\gamma uT<-w/2\\
&\phi^i(\sigma,\sigma-2\gamma uT)=f^i(\sigma)+h^i(-(\sigma-2\gamma uT))-A^i\thickapprox f^i(\sigma)+A^i-A^i=f^i(\sigma)\\
&\left(\frac{\partial}{\partial\omega}-\frac{\partial}{\partial\sigma}\right)\phi^i(\sigma,\omega)=-\left({f^i}'(\sigma)+{h^i}'(-\sigma+2\gamma uT)\right)\thickapprox-{f^i}'(\sigma)\\
\partial\Omega_h: &\thickspace \omega>-w/2\rightarrow \omega+2\gamma uT>w/2\\
&\phi^i(\omega+2\gamma u T,\omega)=f^i(\omega+2\gamma u T)+h^i(-\omega)-A^i\thickapprox A^i+h^i(-\omega)A^i=h^i(\omega)\\
&\left(\frac{\partial}{\partial\omega}-\frac{\partial}{\partial\sigma}\right)\phi^i(\sigma,\omega)=-\left({f^i}'(\omega+2\gamma uT)+{h^i}'(-\omega)\right)\thickapprox-{h^i}'(-\omega)
\end{align*}
Thus, the entire collision initial value problem stated in the Lorentz variables takes the approximate form:
\begin{align}
\frac{\partial^2 {\phi}^i}{\partial\sigma\partial\omega}+\Gamma^i_{j\thickspace k}\frac{\partial{\phi}^j}{\partial\sigma}\frac{\partial \phi^k}{\partial\omega}=\frac{\epsilon}{2}&\frac{\partial V}{\partial {\phi}_i}+\frac{\epsilon}{2}\left[\Gamma^i_{j\thickspace k} \left(\frac{\partial{\phi}^j}{\partial\sigma}\frac{\partial{\phi}^k}{\partial\sigma}+ \frac{\partial{\phi}^j}{\partial\omega}\frac{\partial{\phi}^k}{\partial\omega}\right)-\frac{\partial^2\phi^i}{\partial\sigma^2}-\frac{\partial^2\phi^i}{\partial\omega^2}\right]\\
&\phi^i(\sigma,\omega)  \bigg|_{\partial\Omega_f}  \thickapprox f^i(\sigma)\label{approxbc1}\\
&\frac{\partial}{\partial\sigma}\phi^i(\sigma,\omega)\bigg|_{\partial\Omega_f}\thickapprox{f^i}'(\sigma)\label{approxbc2}\\
&\phi^i(\sigma,\omega)\bigg|_{\partial\Omega_h}\thickapprox h^i(\omega)\label{approxbc3}\\
&\frac{\partial}{\partial\omega}\phi^i(\sigma,\omega)\bigg|_{\partial\Omega_h}\thickapprox -{h^i}'(-\omega)\label{approxbc4}
\end{align}

Now we'll obtain the limiting form of these equations when $\gamma\rightarrow 1$. First we'll turn our attention to the boundary conditions. As $\gamma$ is increased the boundary $\partial\Omega_\gamma$ is pushed along the diagonal with negative slope toward the fourth quadrant. This causes the $\omega$ values of points on $\partial\Omega_f$ to become increasingly negative, and the $\sigma$ values on $\partial\Omega_h$ to become increasingly positive. Consequently, the approximations made in the boundary conditions (\ref{approxbc1}, \ref{approxbc2}, \ref{approxbc3}, \ref{approxbc4}) become ever more accurate. 

This can be seen visually as well. The center of the $f$ soliton occurs, by definition, at the $\omega$-intercept of $\partial\Omega_\gamma$, and the center of the $h$ soliton occurs at the $\sigma$-intercept. As $\partial\Omega_\gamma$ is pushed along the diagonal toward the fourth quadrant the intercepts move away from each other. The distance between the center of each soliton and the place where the boundary is bisected (the endpoint of both half boundaries) increases, resulting in ever more of the $f$ soliton fitting on $\partial\Omega_f$, and the $h$ soliton fitting on $\partial\Omega_h$.\footnote{If one is uncomfortable with this argument for the splitting of the boundary where the initial data is given into two independent pieces, a conformal map can be performed before limit that $\gamma$ goes to infinity is taken. Under a conformal transformation from $\sigma$, and $\omega$ to $\alpha=\tan^{-1}(\sigma)$ and $\beta=\tan^{-1}(\omega)$ the boundary $\partial\Omega_\gamma$ becomes a hyperbola in the $\alpha-\beta$ domain (which is  the square $[-\pi/2,\pi/2]\times[-\pi/2,\pi/2]$). As $\gamma$ is increased the hyperbola is pushed further and further into the lower right of the square, and ultimately becomes the union of the horizontal edge at $\beta=-\pi/2$ ($\sigma$ varying edge), and the vertical at $\sigma=\pi/2$ ($\omega$ varying edge). Though it provides a perhaps a more visually satisfying argument in favor of the split, we do not view the conformal map as necessary.}

Thus, the limiting form of the boundary conditions is obtained by replacing the approximate equalities in \ref{approxbc1}, \ref{approxbc2}, \ref{approxbc3}, and \ref{approxbc4} with equalities. Further, note that the two conditions involving the first derivatives (\ref{approxbc2}, \ref{approxbc4}) no longer contain any additional information than what is captured by the two conditions on $\phi^i$, (\ref{approxbc1}, \ref{approxbc3}). Clearly the limit of the differential equation, \ref{EPSparallelPDE}, is obtained by dropping the $\mathcal{O}(\epsilon)$ term on the righthand side.

At the risk of stating the obvious we'll identify this limiting set of equations with the appropriate collision-- that of the non-Lorentz contracted profiles $f^i$ and $h^i$ each propagating toward one another with speed $u=1$ in the \emph{free} theory. To see why this is the case, take the equation of motion and initial conditions associated with this collision,
\begin{align}
&\Box\phi^i+\Gamma^i_{j\thickspace k}\partial_\mu\phi^j\partial^\mu\phi^k=0\\
&\lim_{t_\rightarrow-\infty}\phi^i(t,x)=f^i(x-t)+h^i(-(x+t))-A^i
\end{align}
and transform to the characteristics, $\xi=x-t$, and $\eta=x+t$. Doing so yields,
\begin{align}
\frac{\partial^2 {\phi}^i}{\partial\xi\partial\eta}&+\Gamma^i_{j\thickspace k}\frac{\partial{\phi}^j}{\partial\xi}\frac{\partial \phi^k}{\partial\eta}=0\label{freePDE}\\
\phi^i(\xi,-\infty)&=f^i(\xi)\label{ic1}\\
\phi^i(\infty,\eta)&=h^i(-\eta)\label{ic2}
\end{align}
Since the righthand side of the resulting differential equation is \emph{identically} zero, we view \emph{this} problem as the limit of the original one (the collision of boosted solitons in the model with nontrivial potential, $V$). So, the approximation to the true solution of the $u<1$ collision problem should be defined by obtaining the solution to the free problem (\ref{freePDE}, subject to \ref{ic1}, and \ref{ic2}), and then evaluating it at the Lorentz variables as opposed to the characteristics. If we denote the solution to the free problem by $\Phi^i(\xi,\eta)$, we mean the approximation for impact velocity $u$ ought to be defined by,
\begin{equation}
\phi^i_{FP}(t,x;u)\equiv \Phi^i(\gamma(x-ut),\gamma(x+ut))\label{FP-definition}
\end{equation}
Turning our attention to $\Phi^i$, we note that it maps $\mathbb{R}^2$ to a submanifold of the field space manifold, $N\subset M$.\footnote{Technically $\Phi^i$ should be viewed as a map from the square $[-s/2,s/2]^2$ in the limit that $s\rightarrow \infty$ boundary conditions given on the edges of the square defined by $(\xi,\eta)=(\xi,-s/2)$ for $\xi\in[-s/2,s/2]$, and $(\xi,\eta)=(s/2,\eta)$ for $\eta\in[-s/2,s/2]$.}
The submanifold is a patch of field space, bounded by four curves. Two of these are simply the original soliton curves (traced out in field space) that we are colliding, since $\Phi^i(\xi,-\infty)=f^i(\xi)$ for $\xi\in\mathbb{R}$, and $\Phi^i(\infty,\eta)=h^i(-\eta)$ for $\eta\in \mathbb{R}$. Significant insight is gained by viewing $\Phi^i$ as the coordinates of two sets of integral curves-- those of one set obtained by varying the first argument and fixing the second, and those of the second set obtained by fixing the first argument and varying the second. Let us name two vector fields these sets of integral curves define as follows,
\begin{align}
U\equiv\frac{\partial}{\partial\xi}=\frac{\partial\Phi^i(\xi,\eta)}{\partial\xi}e_i|_{\Phi(\xi,\eta)}\\
W\equiv\frac{\partial}{\partial\eta}=\frac{\partial\Phi^i(\xi,\eta)}{\partial\eta}e_i|_{\Phi(\xi,\eta)}\
\end{align}
where we've expanded in the coordinate basis $\{e_i\}=\{ \frac{\partial}{\partial\Phi^i}\}$. Expressed in terms of the vector fields $U$ and $W$, the boundary conditions for $\Phi^i$ simply indicate that the vector fields at the relevant two edges of the submanifold line up with the tangent vectors to the two original soliton curves. In an effort to minimize confusion with the negative signs, we explicitly point out where the vacuum values are in the submanifold, parameterized by $\xi$ and $\eta$: $\Phi(\infty,-\infty)=A$, $\Phi(-\infty,-\infty)=B$, and $\Phi(\infty,\infty)=C$.

The differential equation takes the form,
\begin{align}
0=\frac{\partial^2\Phi^i}{\partial\eta\partial\xi}+\Gamma^i_{j\thickspace k}\frac{\partial\Phi^j}{\partial\xi}\frac{\partial\Phi^k}{\partial\eta}&=\frac{\partial U^i}{\partial \eta}+\Gamma^i_{j\thickspace k}U^j\frac{\partial\Phi^k}{\partial\eta}\\
&=\frac{\partial\Phi^\ell}{\partial\eta}\frac{\partial}{\partial\Phi^\ell} U^i+\frac{\partial\Phi^k}{\partial\eta}U^j\Gamma^i_{j\thickspace k}\\
&=\frac{\partial\Phi^\ell}{\partial\eta}\left(e_\ell[U^i]+U^j\Gamma^i_{j\thickspace\ell}\right)\\
&=W^\ell\left(e_\ell[U^i]+U^j\Gamma^i_{j,\ell}\right)
\end{align}
Since the equality holds for each component we have,
\begin{align}
0&=W^\ell\left(e_\ell[U^i]e_i+U^j\Gamma^i_{j\thickspace\ell}e_i\right)\\
&=W^\ell\left(\nabla_{e_\ell}(U^i e_i)\right)\\
&=\left(\nabla_{W^\ell e_\ell}(U^i e_i)\right)=\nabla_W U\label{parallel-trans}
\end{align}
Similarly, we obtain $\nabla_U W=0$ by the analogous series of steps (when $\xi$ and $\eta$ are swapped, since \ref{freePDE} is symmetric under exchange of these). We thus arrive at the geometrical description of bubble collisions in the ultrarelativistic limit: the resulting field profiles post-collision are determined by the mutual  parallel transport of each soliton's tangent vector field along that of the other soliton. That is, the tangent vector fields of the soliton profiles are parallel transported along each other everywhere in $N$.

\begin{figure}
\includegraphics[width=\textwidth]{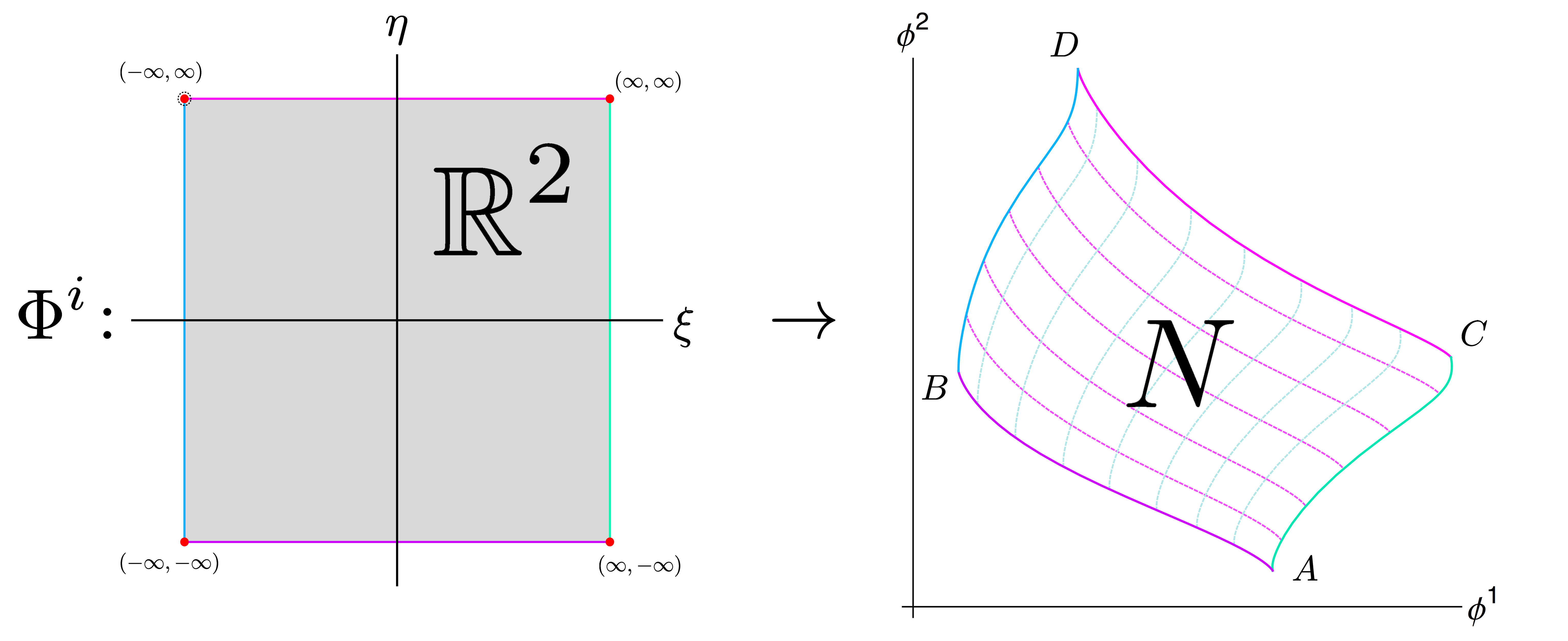}
\caption{The soliton collision initial value problem, expressed in the Lorentz variables takes the form \ref{freePDE}, with \ref{ic1}, \ref{ic2} when the limit that the impact velocity goes to $1$ is taken. The solution to this limiting set of equations, denoted by $\Phi^i$ maps $\mathbb{R}^2$ to a submanifold, $N$, of the field space $M$. Since the map is smooth the image of the $\xi$ coordinate lines and the $\eta$ coordinate lines are the integral curves of two vector fields. The partial differential equation \ref{freePDE} indicates these two vector fields are parallel transported along one another everywhere in $N$. The initial conditions require that $\Phi(\xi,\eta)$ go to $f(\xi)$ as $\eta\rightarrow -\infty$, and go to $h(-\eta)$ as $\xi\rightarrow 1$. This is shown in the cartoon/schematic illustration above, where $\mathbb{R}^2$ is drawn as a (finite) square. The purple horizontal line in the lower half of the $\xi$-$\eta$ plane is mapped to the purple curve with endpoints at $B$ and $A$ in the $\{\phi^i\}$ coordinate plane($f$ soliton), and the green vertical line in the right half plane is mapped to the green curve with endpoints $C$ and $A$ ($h$ soliton). The remaining two curves that form the rest of the boundary of $N$ are shown in blue and pink. They are the images of the $\xi$ coordinate line at $\eta\rightarrow \infty$ and $\eta$ coordinate line at $\xi\rightarrow- \infty$, so are in a sense the curves obtained by completing the transport of $h$ along $f$ and $f$ along $h$. At sufficiently high impact velocity the field in the collision region takes on value $D$, and the outgoing walls interpolate between $D$ and the original bubble vacua, $B$ to the right and $C$ to the left. For such a collision the parametric plot of the two walls differs negligibly from the prediction via parallel transport-- the blue and pink curves.}  
\label{submani-Fig}
\end{figure}

The remaining two curves that together with $\Phi^i(\xi,-\infty)$, and $\Phi^i(\infty,\eta)$ form the boundary of $N$ are simply $\Phi^i(\xi,\infty)$ and $\Phi^i(-\infty,\eta)$. The first of these, $\Phi^i(\xi,\infty)$, goes between $C^i$ when $\xi=\infty$, and the point $\Phi^i(-\infty,\infty)\equiv D^i$. The second, $\Phi^i(-\infty,\eta)$, has endpoint $B^i$ when $\eta=\infty$, and the other at $D^i$ as well, when $\eta=-\infty$. This is shown schematically in Figure \ref{submani-Fig}.

This result agrees with heuristic expectations motivated by the flat field space limit. Namely, note that in the flat limit, a right moving soliton that interpolates between the parent vacuum $A$  and a local minimum $B$ leaves in its wake (to the left of the soliton's transition wall) a field value shifted by $\Delta_L$ = $B - A$, while a left moving soliton that interpolates between the parent vacuum $A$  and a local minimum $C$ leaves in its wake (to the right of the soliton's transition wall) a field value shifted by $\Delta_R$ = $C - A$. Thus, after free-passage collision, the collision region -- which, by definition is to the left of the right-mover and to the right of the left-mover -- is shifted by $\Delta_L + \Delta_R$ (which equals $B + C - 2 A$). In the case of curved field space, we divide the field shifts, both $\Delta_L$ and $\Delta_R$, into infinitesimals, which geometrically are the tangent vector fields of the soliton field profiles. Each such infinitesimal leaves in its immediate wake a field whose value is parallel transported along the infinitesimal shift vector, thus resulting in the geometrical picture we've described. When all tangent vectors of nontrivial magnitude have been mutually transported, they leave a widening interior of field in $\phi=D$. 

This type of reasoning indicates that \ref{parallel-trans} is the simplest partial differential equation that reduces to free passage in the flat field space limit. Namely, in the flat limit, the infinitesimal description of free passage is clearly the requirement that the vanishing of the directional derivatives of $U$ with respect to $W$ and $W$ with respect to $U$. The covariant version of these statements is just \ref{parallel-trans}. This heuristic argument is suggestive but not sufficient since it is insensitive to any terms in the limiting form of the partial differential equation that vanish in the flat field space limit but which could nonetheless be present in the curved case. The analysis we've performed so far, together with that in the following section verifies that there are no such terms.

It is worth confirming that our free passage field configuration, \ref{FP-definition}, does indeed have the qualitative features outlined above. First note that the configuration correctly approaches the $B$ vacuum asymptotically to the left, and the $C$ vacuum to the right for any finite time $t$, since this amounts to evaluating $\Phi$ at $(\gamma(x-ut),\gamma(x+ut))\rightarrow(-\infty,-\infty)$, and $(\infty,\infty)$, respectively. At a fixed time \emph{before} collision, any time $t\lesssim -w/2\gamma u$, the free passage field differs from $B$ vacuum by an insignificant amount at $x<ut-w/2\gamma$ since we'd effectively be evaluating $\Phi^i$ in \ref{FP-definition} at $(-\infty,-\infty)$. As we march rightward the first argument increases, and reaches zero at $x=ut$ while second argument remains essentially unchanged. $\Phi(0,-\infty)$ is simply the center of the $f$ soliton. So, as one moves between the positions $-ut-w/2\gamma u$, and $-ut+w/2\gamma u$ in free passage field configuration they run through the $f$ soliton's field configuration. If they continue moving rightward they'll reach a stretch of $x$ values where both the arguments of $\Phi^i$ in \ref{FP-definition} are effectively negative infinity, and so the $A$ vacuum is measured. 

If we continue on rightward the analogous procedure 
leads us to realize that the free passage field configuration interpolates between the $A$ and $C$ vacua by the (reflected) $h$ soliton, centered at $-ut$. So, pre-collision the spatial profile of the free passage field configuration looks like the usual linear superposition: a nearly homogeneous interior of diminishing size in the parent vacuum, separated from the bubble vacua by the relevant solitons, whose centers lie at $ut$, and $-ut$.

The same line of reasoning can be used to deduce that post-collision, any time $t\gtrsim w/2\gamma u$, the free passage configuration again consists of three approximately homogeneous regions: a widening interior, or ``collision region" with field value $\thickapprox D^i$, separated from regions of original bubble vacua on either side, by walls whose centers follow the same trajectories $x=\pm ut$. The shapes however of the spatial profiles of the field components across the walls are \emph{not} in general the same as those of the incoming solitons. A parametric plot of the free passage configuration at a given time (with the spatial variable as the parameter) in the $\{\phi^i\}$ coordinate plane would consist of the composition of a curve that interpolates between $B$ and $D$, together with the one between $C$ and $D$. These would be nearly identical those obtained by completing parallel transport, and approaches the union of these two curves, $\Phi^i(-\infty,\infty)$, and $\Phi^i(\infty,\infty)$ asymptotically as $t\rightarrow\infty$. 

We claim that there always exists an impact velocity sufficiently close to the speed of light such that the actual solution to \ref{collisionIVP} is well approximated by the above free passage evolution throughout the entirety of the collision-- i.e. for longer than the amount of time it would take for the incoming Lorentz contracted walls to fully pass through each other. We prove this in the following section.

\section{Proof of Realization of Free Passage}
The solution to the parallel transport problem, $\Phi^i$, and the free passage evolution function defined from it, $\phi^i_{FP}$, is, of course, only useful in predicting the outcome of a particular collision if deviations from $\phi^i_{FP}$ remain sufficiently small throughout the entirety of the collision (or longer). As we've mentioned previously, the amount of time it takes the solitons to fully pass through each other is $w/\gamma u$, and since we've chosen to set our $t=0$ at the middle of the collision we're interested in the time period, $t\in[-w/2\gamma u, w/2\gamma u]$. 

We begin by expanding the actual solution (to \ref{collisionIVP}) about the free passage configuration,
\begin{equation}
\phi^i(t,x;u)=\phi^i_{FP}(t,x;u)+\psi^i(t,x;u)
\label{expansion}
\end{equation}
Simply substituting \ref{expansion} into the equation of motion and expanding in powers of $\psi$ yields,
\begin{align}
\Box\psi^i&=-\frac{\partial V}{\partial\phi_i}\bigg|_{\phi_{FP}}\negthickspace\negthickspace\negthickspace\negthickspace-\frac{\partial^2 V}{\partial\phi^\ell\partial\phi_i}\bigg|_{\phi_{FP}} \negthickspace\negthickspace \psi^\ell +\mathcal{O}(\psi^2)-\left[\Box\phi^i_{FP}+\Gamma^i_{j k}|_{\phi_{FP}} \partial_\mu\phi^j_{FP} \partial^\mu\phi^k_{FP} \right]\\
&-2\Gamma^i_{j k}|_{\phi_{FP}}\partial_\mu\phi_{FP}^j \partial^\mu\psi^k-\Gamma^i_{j k,\ell}|_{\phi_{FP}}\psi^\ell \partial_\mu\phi^j_{FP} \partial^\mu\phi^k_{FP}+\mathcal{O}(\psi(\partial\psi))+\mathcal{O}((\partial\psi)^2)\\
&=-\frac{\partial V}{\partial\phi_i}\bigg|_{\phi_{FP}}\negthickspace\negthickspace\negthickspace\negthickspace-\left[\Box\phi^i_{FP}+\Gamma^i_{j k}|_{\phi_{FP}} \partial_\mu\phi^j_{FP} \partial^\mu\phi^k_{FP} \right]+\mathcal{O}(\psi)
\label{GreenFN}
\end{align}
We can write an implicit expression for $\psi^i(t,x)$ by integrating the right hand side of \ref{GreenFN} as follows, 
\begin{equation}
\psi^i(t,x)=\int_{-w/2\gamma u}^{t}dt'\int_{x-t'}^{x+t'}dx'G^i(t',x')
\label{psiInt}
\end{equation}
We now truncate at zeroth order in $\psi$, and bound the above integral. The term in the square brackets in \ref{GreenFN} is,

\begin{align}
-\left[\Box\phi^i_{FP}+\Gamma^i_{j k}|_{\phi_{FP}} \partial_\mu \phi^j_{FP} \partial^\mu\phi^k_{FP} \right]&=
4(1-\epsilon)\gamma^2\left[\frac{\partial^2 \Phi^i}{\partial\sigma\partial\omega}+\Gamma^i_{j\thickspace k}\frac{\partial\Phi^j}{\partial\sigma}\frac{\partial \Phi^k}{\partial\omega} \right]\label{mismatch}\\
&-2\gamma^2\epsilon\left[\frac{\partial^2\Phi^i}{\partial\sigma^2}+\frac{\partial^2\Phi^i}{\partial\omega^2}+\Gamma^i_{j\thickspace k} \left(\frac{\partial\Phi^j}{\partial\sigma}\frac{\partial\Phi^k}{\partial\sigma}+ \frac{\partial\Phi^j}{\partial\omega}\frac{\partial\Phi^k}{\partial\omega}\right)\right]\nonumber
\end{align}
where we've expressed the operators $\Box$ and $\partial_\mu$ in terms of the the Lorentz variables, and retained terms up to first order in $\epsilon$. This is identical to the step we took at the outset to obtain \ref{LorentzVarsEOM}. Note that the first term in square brackets on the righthand side of $\ref{mismatch}$ is, by definition, zero. The second term, however, does not vanish. It results from the mismatch between the Lorentz variables and the characteristics. The nonvanishing piece can be expressed in terms of the vector fields $U$ and $W$ as follows,
\begin{align}
-2\gamma^2\epsilon\left[\frac{\partial^2\Phi^i}{\partial\sigma^2}+\Gamma^i_{j\thickspace k} \frac{\partial\Phi^j}{\partial\sigma}\frac{\partial\Phi^k}{\partial\sigma}+\frac{\partial^2\Phi^i}{\partial\omega^2} +\Gamma^i_{j\thickspace k} \frac{\partial\Phi^j}{\partial\omega}\frac{\partial\Phi^k}{\partial\omega}\right]=-2\gamma^2\epsilon\left[\left(\nabla_U U\right)^i+\left(\nabla_W W\right)^i\right]
\end{align}

A bound on the magnitude of $\psi^i$ can now be computed straightforwardly, 
\begin{align}
|\psi^i(t,x)|&\leq \int_{-w/2\gamma}^{t}dt'\int_{x-t'}^{x+t'}dx' |G^i(t',x')|\\
&\leq \int_{-w/2\gamma}^{t}dt'\int_{x-t'}^{x+t'}dx' \bigg|\frac{\partial V}{\partial \phi_i}\bigg|+2\gamma^2\epsilon\left|\left(\nabla_U U\right)^i\right|+2\gamma^2\epsilon\left|\left(\nabla_W W\right)^i\right|
\end{align}
where the terms involving the vector fields are evaluated at $(\gamma(x'-ut'),\gamma(x'+ut'))$. Now, we're only interested in the deviation at points $x$ in the collision region (outside of here the field persists very nearly equal to the bubble vacuum field values), and times $t\in[-w/2\gamma u,+w/2\gamma u]$. For this time period the collision region is always contained within $[-w/\gamma,+w/\gamma]$. This means the $x'$ interval we need to integrate over in our expression for $\psi^i(t,x)$ is always contained within $[-3w/2\gamma,+3w/2\gamma]$. So we can write,
\begin{align}
|\psi^i&(t,x)|\leq \int_{-w/2\gamma}^{t}dt'\int_{-3w/2\gamma}^{3w/2\gamma}dx'|G^i(t',x')|\\
&\leq\left\{ \sup_{\phi\in N}\left(\bigg|\frac{\partial V}{\partial \phi_i}\bigg|\right)+2\gamma^2\epsilon \negthickspace\negthickspace \sup_{(\sigma,\omega)\in\mathbb{R}^2}\negthickspace \left(\left|\left(\nabla_U U(\sigma,\omega)\right)^i\right|+\left|\left(\nabla_W W(\sigma,\omega)\right)^i\right|\right)\right\} \negthickspace \int_{-w/2\gamma u}^{t}\negthickspace\negthickspace\negthickspace\negthickspace   dt'\int_{-3w/2\gamma}^{3 w/2\gamma}\negthickspace\negthickspace\negthickspace\negthickspace dx'\\
&=(k_1^i+2\gamma^2\epsilon k_2^i)\frac{w(t/T+1/2)}{\gamma u}\frac{3 w}{\gamma}=\left(\frac{k_1^i}{\gamma^2}+2\epsilon k_2^i\right)3w^2=\left(2\epsilon k_1^i+2\epsilon k_2^i\right)3w^2
\end{align}
So,
\begin{equation}
|\psi^i(t,x)|\leq k^i\epsilon
\end{equation}
where the positive constants,
\begin{align}
k^i\equiv 6w^2\left\{ \sup_{\phi\in N}\left(\bigg|\frac{\partial V}{\partial \phi_i}\bigg|\right)+2\gamma^2\epsilon \negthickspace\negthickspace \sup_{(\sigma,\omega)\in\mathbb{R}^2}\negthickspace \left(\left|\left(\nabla_U U(\sigma,\omega)\right)^i\right|+\left|\left(\nabla_W W(\sigma,\omega)\right)^i\right|\right)\right\},
\end{align}
are finite due to the smoothness of the potential and the field space manifold.

Since the difference in the coordinates of the true field configuration and free passage configuration can be made arbitrarily small, we conclude that the post collision field (for any two solitons in any curved multi-scalar field theory) successfully realizes the late-time free passage field configuration, provided the impact velocity was sufficiently relativistic. The threshold above which the impact velocity ought to be is dependent on both the model and the choice of the two colliding solitons. This threshold can be estimated by requiring that the distance in field space between the free passage field (say for the center of the collision region) at time $t$, and the free passage plus deviation location be much much smaller than the length of the path the observer at the center of the collision region has through field space from the parent vacuum until time $t$. Since the walls of bubbles nucleated via Coleman-De Luccia tunneling accelerate as they move outwards, we expect our parallel transport procedure to be a useful means of predicting the field configuration following the collision of two bubbles, provided they were nucleated sufficiently far apart (and they're radii upon nucleation is sufficiently small compared to the separation distance such that high enough impact velocity is reached upon collision). 

\section{Numerical Simulations}

We simulated soliton collisions at a variety of impact velocities in three different models with actions of the form \ref{Action}. Each model featured a different two dimensional curved field space. We reiterate that the field space is curved in the sense that the matrix of $\{\phi^i\}$ dependent functions, $g^{ij}$, in the noncanonical kinetic term in the Lagrangian is the coordinate representation of the metric on a curved manifold (clearly the field components $\{\phi^i\}$ are identified as coordinates on the manifold). The particular manifolds we considered were the sphere, the ring torus, and the ``teardrop"-- our own creation named for obvious reasons.

For both the sphere and teardrop we used the polar angle and azimuthal angle as our two coordinates. For the torus we used the angles about the major axis, and the minor axis. To minimize the possibility for confusion we adopt standard naming conventions used for these coordinate systems, and refer to the field components ($\phi^1,\phi^2$) as ($\theta, \phi$) for the sphere and teardrop, and as ($u, v$) for the torus. For the explicit form of the metric components, as well as vacuum locations refer to Table, \ref{Models-Table}.

\begin{table}
\caption{Metric Components and Vacuum Locations}
\centering
\begin{tabular}{| l | l| c |c |c | c|}
\hline\hline
Geometry&{$g^{ij}$}&\multicolumn{3}{|c|}{Parent and Bubble Vacua}& Free Passage \\
& & A& B& C& D\\
\hline
Sphere & $g_{\theta\theta}=1  $&    $(\frac{\pi}{12}, \frac{2\pi}{15}  )$& $(\frac{5\pi}{12},\frac{2\pi}{11})$ & $(\frac{3\pi}{12},\frac{6\pi}{13} )$& (1.77,1.08) \\
          &$g_{\phi\phi}=  \sin^2(\theta)$& & & & \\
          \hline
Teardrop & $g_{\theta\theta}= \cos^2(\theta)+\left[\frac{\sin(\theta)}{2}+(\theta-\pi) \right]^2$&$ (\frac{7 \pi}{60} ,\frac{\pi}{13})$ &$(\frac{118 \pi}{327},-\frac{\pi}{13})$ &$(\frac{73\pi}{327},\frac{35\pi}{109})$&(1.52,0.33)\\
& $g_{\phi\phi}=  \sin^2(\theta)$& & &&  \\
\hline
Torus &$g_{uu}= (1+.7\cos(v))^2$& $ (\frac{145 \pi}{654} ,\frac{-20\pi}{109})$ &$(-\frac{145 \pi}{654},\frac{20\pi}{109})$ &$(-\frac{35\pi}{654},\frac{235\pi}{654})$&(1.36,0.09) \\
 & $g_{vv}= .7^2$& & &  &\\
          \hline
          \end{tabular}
\label{Models-Table}
\end{table}

We numerically approximated the solutions to the initial value problem \ref{collisionIVP} associated with the collision of two non-identical solitons in the given theory, as well as solutions to the mutual parallel transport of two tangent vector fields problem, \ref{freePDE}-\ref{ic2}, using Mathematica's finite difference partial differential equation solver, NDSolve. The potential was engineered to have three degenerate vacua with generic looking wells and barriers by using the product of two trigonometric functions, and then isolating only three minima by multiplying by a superposition of hyperbolic tangents which served as smooth approximations to characteristic functions and hat functions. For the explicit form of the three potentials see Table \ref{potentials-Table}. 

We wanted the potential to be flat outside the neighborhood immediately surrounding the three vacua so as to minimize the influence of the potential on field dynamics, both throughout the collision and after, so that the free passage behavior could feasibly be extracted. Though we absolutely assert that parallel transport is generic (there always exists a speed high enough such that it is fully realized), we wanted to design a nontrivial scenario where the boost needed was small enough, and so the grid size large enough, that we'd have a hope of resolving this in Mathematica, and on a desktop computer.

\begin{figure}[b!]
\includegraphics[width=1.1\textwidth]{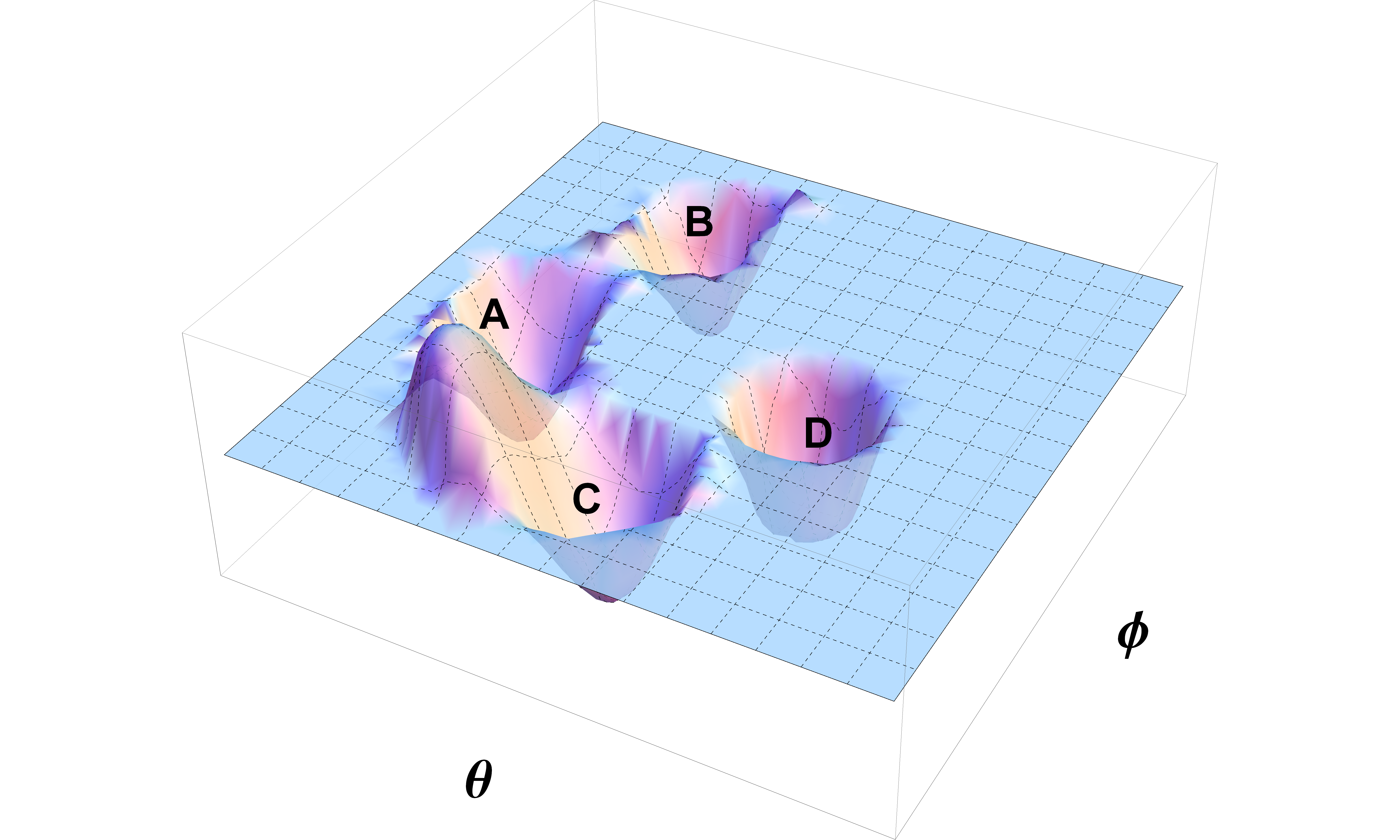}
\caption{Plot of the potential for the teardrop model. Note the cylindrical well carved out of the plateau at $D$. This addition does not change the solitons $f$, and $h$.}
\label{teardrop-potential}
\end{figure}

There is a final step to designing a potential that enables us to extract the free passage dynamics-- the placement of a fourth degenerate vacuum at the free passage location, $D$, which of course is not known a priori. Had the potential been left as a plateau outside the three vacua the post collision field dynamics would be tainted by the pressure gradient across outgoing walls resulting from the energy density in the collision region differing from that in the surrounding region (which is still simply that of the degenerate bubble vacua). In order to prolong the amount of time after which free passage would remain a good approximation, without unduly biasing the field toward the free passage field location we carved a cylindrical well out of the plateau at the free passage field location, for each model. Clearly then, the parallel transport solution for each geometry was obtained before any collisions were simulated, so that each of their potentials could be modified in the manner described. 

Note that the parallel transport solution, $\Phi$, is by definition independent of the potential provided that the soliton curves between the parent and two bubble vacua remain unchanged (since these are the boundary conditions in the parallel transport problem). Clearly the potential in the neighborhood of the three original vacua is unaffected by the addition of the narrow cylindrical well placed out on the plateau away from the original three vacua. A plot of the potential in the teardrop model is included as an example in Figure \ref{teardrop-potential}, and the explicit form of the potentials used for all three geometries can be found in Table \ref{potentials-Table}. \footnote{where,
\begin{align*}
 &\chi_{sphere}(\theta,\phi)= (1 - \tanh((\theta - .25\phi - 5\pi/12 - .2) 30)))(1 - \tanh((\phi + 1.5 \theta - 3.2) 30))/4\\
& \chi_{tear}(\theta,\phi)=  \frac{1}{2^6} (1 + \tanh[40 (-0.05 + (-1.56 + u)^2 + (0.06 + v)^2)]) (1 + 
   \tanh[20 (-0.5 + (-2.42 + 2 u)^2 + (-2.3 + 2 v)^2)]) \\
 & \qquad *(1 +   \tanh[20 (-0.75 + (-0.07 + 2 u)^2 + (-2 + 2 v)^2)]) (1 +  \tanh[10 (-3.2 + (-3.87 + 2 u)^2 + 1/2 (-1.9 + 2 v)^2)]) \\
 &\qquad*(1 + \tanh[20 (-1 + (-2.27 + 2 u)^2 + (-1 + 2 v)^2)]) (1 + \tanh[10 (2.1 - (-1.77 + 2 u)^2 - 1/2 (-0.8 + 2 v)^2)])\\
&\chi_{torus}(u,v)= \frac{1}{2^6}(\tanh(10 (2.1 - (v - .8)^2/2 - (u - .2)^2)) + 1)*  (1 - \tanh(30 ((.15 v + u) - .9)))\\
&\qquad*(\tanh(10 ((v - 1.9)^2/2 + (u - 2.3)^2 - 3.2)) + 1)* (\tanh(20 ((v - 1)^2 + (u - .7)^2 - 1.2)) + 1)\\
&\qquad* (\tanh(20 ((v - 2.3)^2 + (u - .85)^2 - .5)) + 1) *(\tanh(20 ((v - 2)^2 + (u + 1.5)^2 - .75)) + 1) 
\end{align*} }

\begin{table}[h!]
\caption{Below is the explicit form of the potentials we used in our simulations of solitons collisions for each of the field space geometries we considered.}
\centering
\begin{tabular}{|l| l |}
\hline
Model & Potential\\
\hline
\hline
Sphere& $V(\theta,\phi)=V_0\sin(6 \theta)\sin\left(\negthickspace \frac{(3 \theta - 4\pi) \phi}{\pi}\right)\negthickspace *\negthickspace \chi_{sphere}(\theta,\phi)-\negthickspace \frac{V_0}{2}\tanh(40( (\theta-D_\theta)^2\negthickspace+\negthickspace (\phi-\negthickspace D_\phi \negthickspace -\negthickspace .3)^2))\negthickspace +2.5)$\\
&$\qquad+\frac{V_0}{2}\tanh(40 ((\theta-D_\theta)^2+(\phi-D_\phi-.3)^2)-2.5)$\\
\hline
Teardrop& $V(\theta,\phi)=V_0\sin(4\phi-1.2\theta+.3\pi)\sin(6\theta-1.5\pi+1.8 \phi)*\chi_{tear}(\theta,\phi)$\\
&$\qquad-\frac{V_0}{2}\left[\tanh( 2.7-  60 (((\theta - D_\theta)^2 + (\phi - D_{\phi})^2))) +1\right]$\\
\hline
Torus& $V(u,v)=V_0\sin(2 v - .6 u) \sin(3 u + .9v)*\chi_{torus}(u,v)$\\
&$\qquad-\frac{V_0}{2}\left[\tanh( 3-  30((u -D_u)^2 + (v - D_v)^2))) +1\right]$\\
\hline
  \hline
          \end{tabular}
\label{potentials-Table}
\end{table}

Lastly it is necessary to discuss how the initial conditions and boundary conditions were formulated. Both are defined in terms of the components of the two solitons we are colliding. Solitons are, by definition, static solutions to the equations of motion \ref{curved-EOM} that approach two distinct (obviously degenerate) minima of the potential asymptotically. In a multi-scalar field theory solitons are unique to the vacua they interpolate between, and furthermore are the minimum energy field configurations that satisfy the given pair of boundary conditions. Since the coupled ordinary differential equations that define the solitons are nonlinear, analytic solutions generally cannot be found. However, if an initial profile that satisfies the boundary conditions is evolved in time by the equations of motion plus a damping term, the profile ultimately settles down to the soliton, provided the initial guess was sufficiently close to the true soliton and the damping coefficient was not too large. We performed this relaxation procedure numerically, once again with NDSolve in Mathematica.\footnote{It is important to mention how the initial guesses for the soliton components in relaxation were chosen. Since the (true) soliton is defined by both the geometry and the potential, we sought to allow both to play a role in our guesses. For a given pair of vacua we first parameterized the geodesic connecting them by writing one field component in terms of the other (for instance, in the case of the sphere the geodesics were great circles and the polar angle was parameterized in terms of the azimuthal). The potential was then evaluated along the geodesic, and the resulting function was approximated as a double well potential, which has a single free parameter after the distance between the minima is fixed. This parameter was tuned such that the approximate potential not only qualitatively resembled the true one along the geodesic, but also so that their integrals of the inverse square root of the difference between the vacuum value, $-V_0$, and the potential, between the minima were nearly identical for the. For example, for the sphere initial guess we'd compute
\begin{equation}
\int_{\phi_A}^{\phi^B}d\phi/\sqrt{V_{sphere}(c_{geo,{AB}}(\phi),\phi)+V_0}
\end{equation} 
numerically and tune the double well potential's curvature parameter until its integral matched this.
The double well approximation then provides us with an initial guess, a (scaled and shifted) hyperbolic tangent, for one of the two soliton components-- that which the geodesic is parameterized in terms of. To obtain a guess for the remaining component the expression for the geodesic was simply evaluated at the guess function that was just obtained for the former component-- resulting in a spatial profile.} 

Note that analytic expressions for the four soliton components, $f^i$, and $h^i$, were needed in order for simulations of the collision to be feasible. At such small grid spacing time evolving initial conditions constructed out of the interpolating functions relaxation yielded was not possible. So the final step was to engineer analytic expressions that approximated each of the soliton components from relaxation (four total, $f^1$, $f^2$, $h^1$, $h^2$). All were modifications of (scaled and shifted) hyperbolic tangents, typically with the addition of small gaussians and nonlinear terms in the argument of the hyperbolic tangent. 

\begin{figure}
{\includegraphics[width=.35\textwidth]{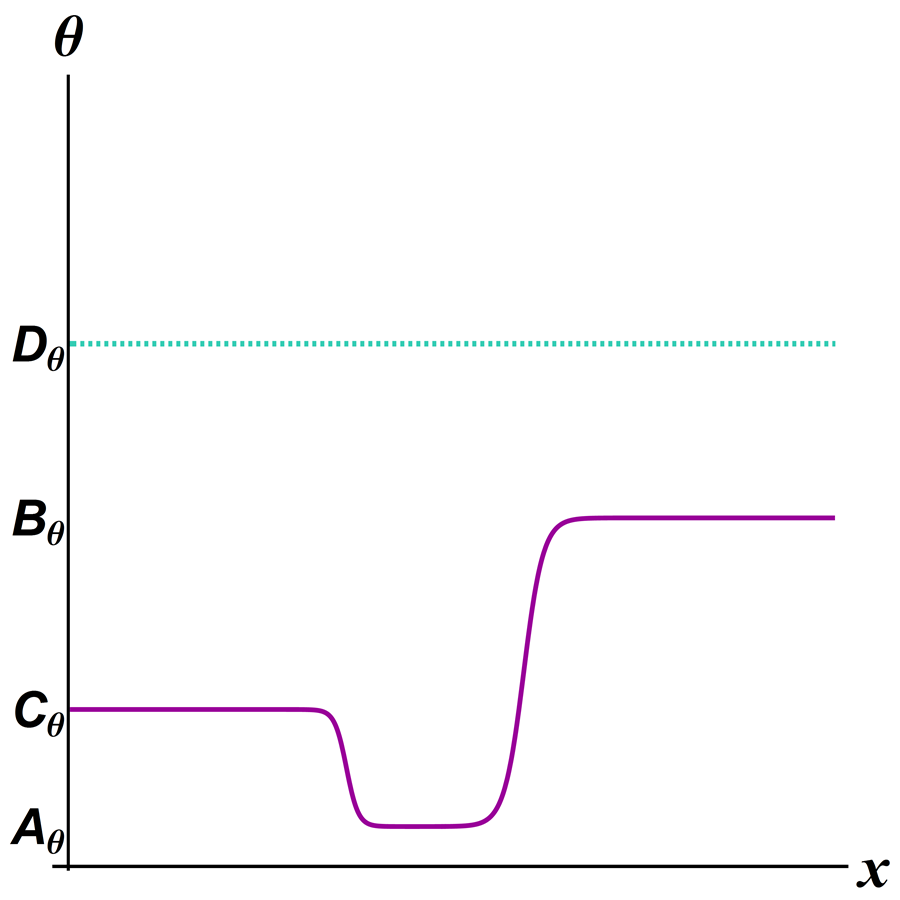}\includegraphics[width=.35\textwidth]{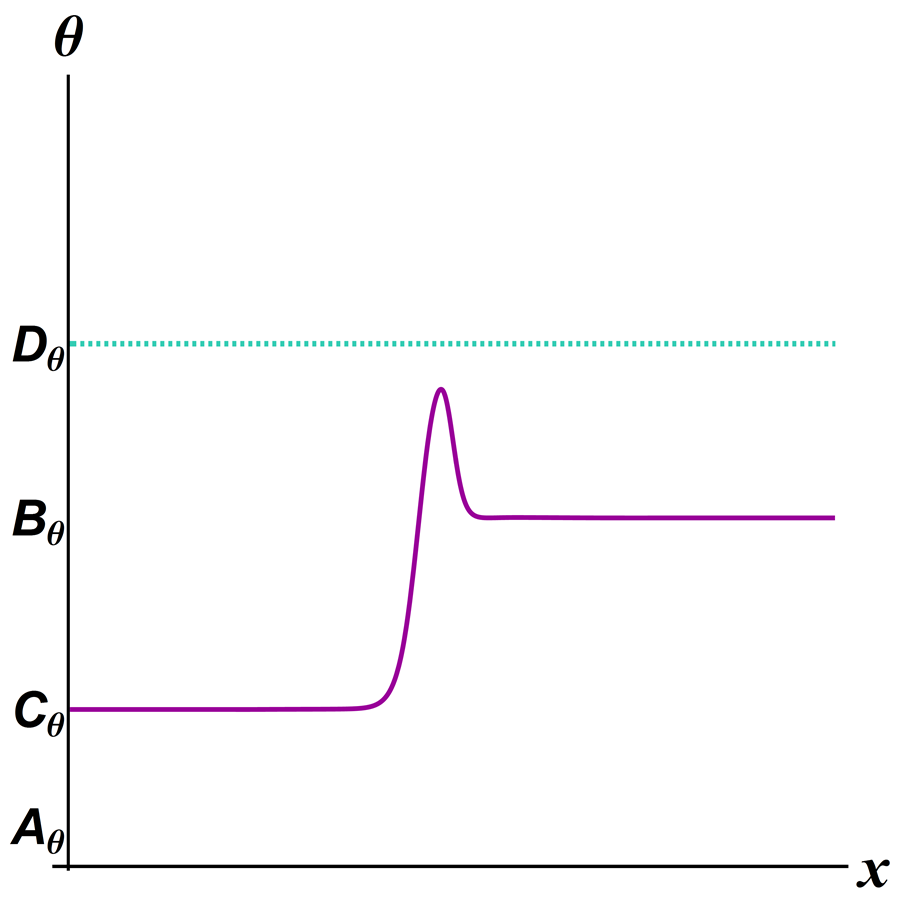}\includegraphics[width=.35\textwidth]{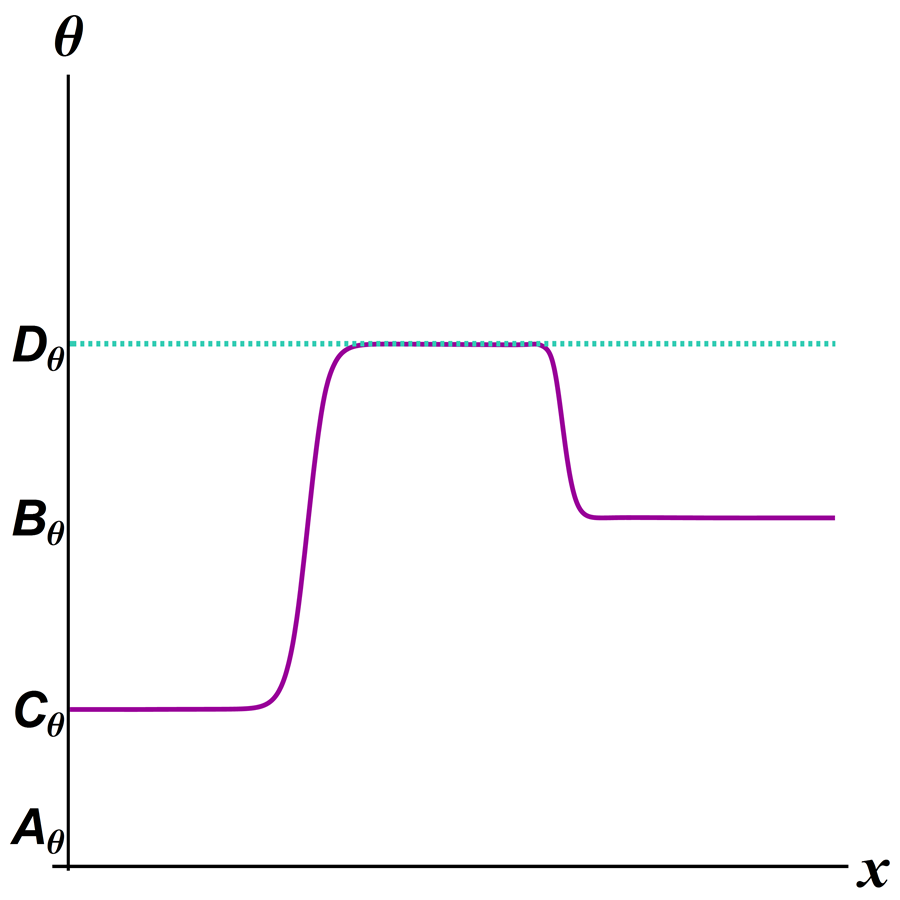}}\\
{\includegraphics[width=.35\textwidth]{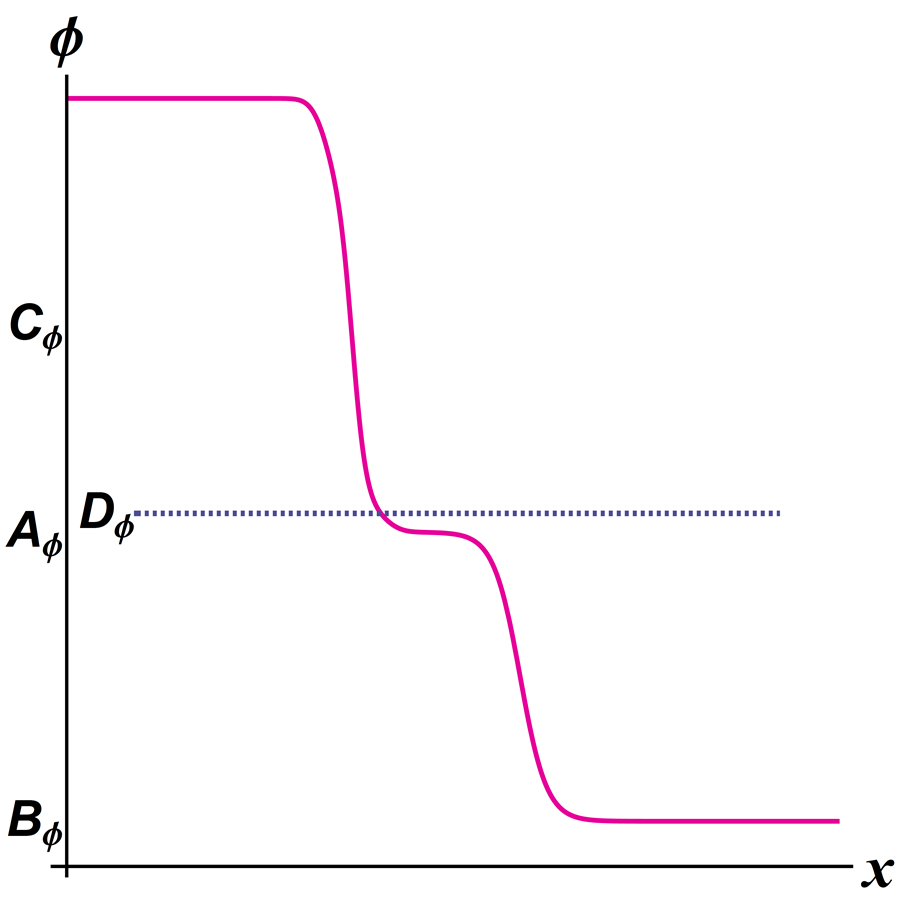}\includegraphics[width=.35\textwidth]{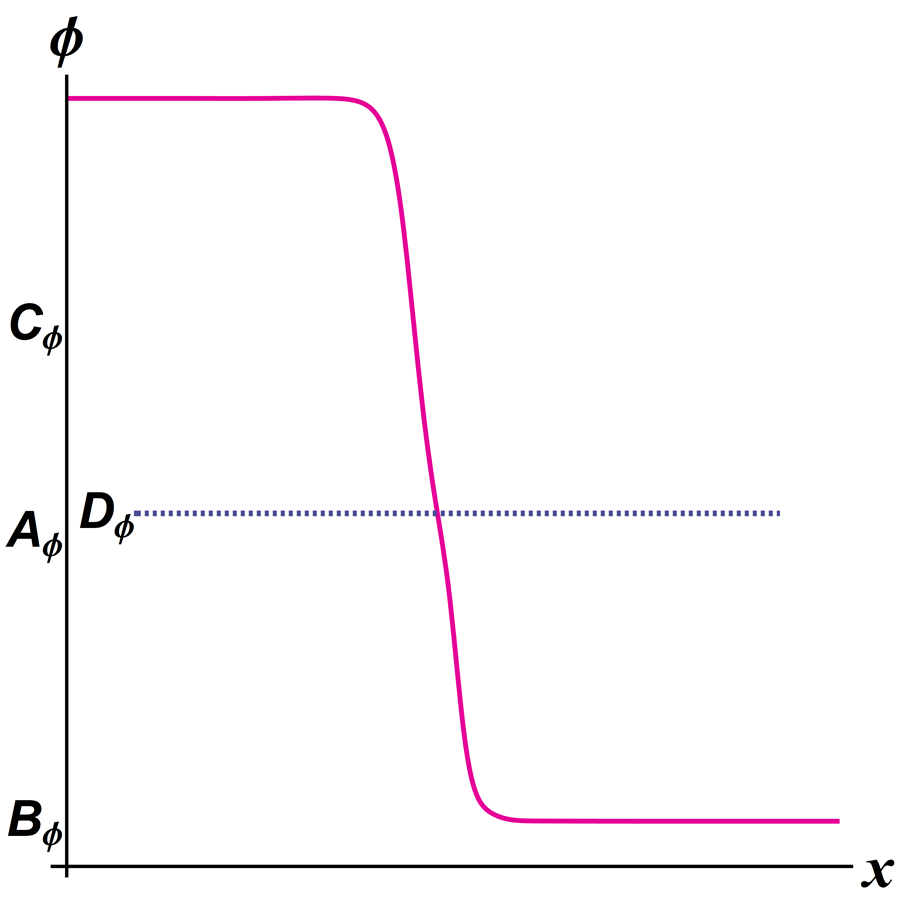}\includegraphics[width=.35\textwidth]{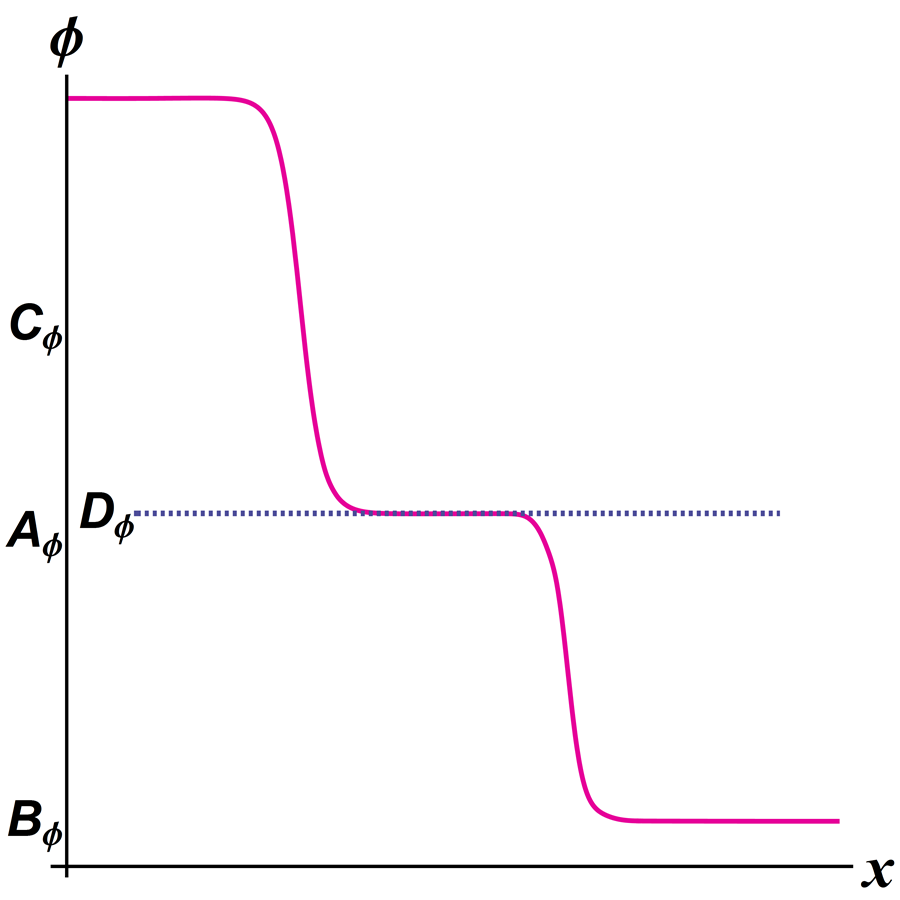}}
\caption{Here we show snapshots of each field component of the configuration during a collision simulated in the teardrop model at impact velocity $u=.995$ ($\theta$ component is on the top row, and $\phi$ on the bottom row). The prediction of each component's value inside the collision region obtained by parallel transport, $D_i$ are indicated by the dashed teal and purple horizontal lines for $\theta$, and $\phi$, respectively. Note both the homogeneity of the field in the collision region, and its extraordinarily strong agreement with the free passage prediction.}
\label{snapshots-tear}
\end{figure}

\begin{figure}
{\includegraphics[width=.5\textwidth]{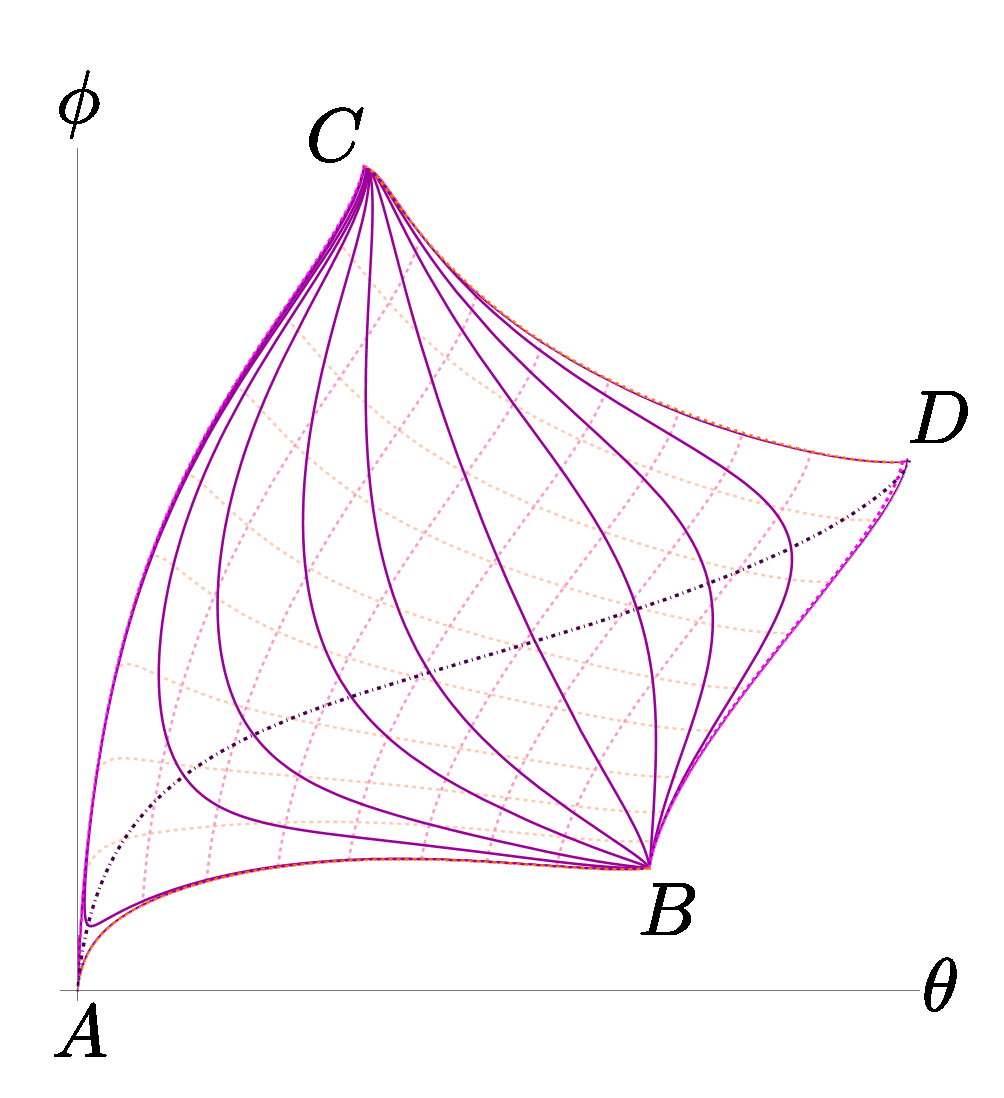}\includegraphics[width=.55\textwidth]{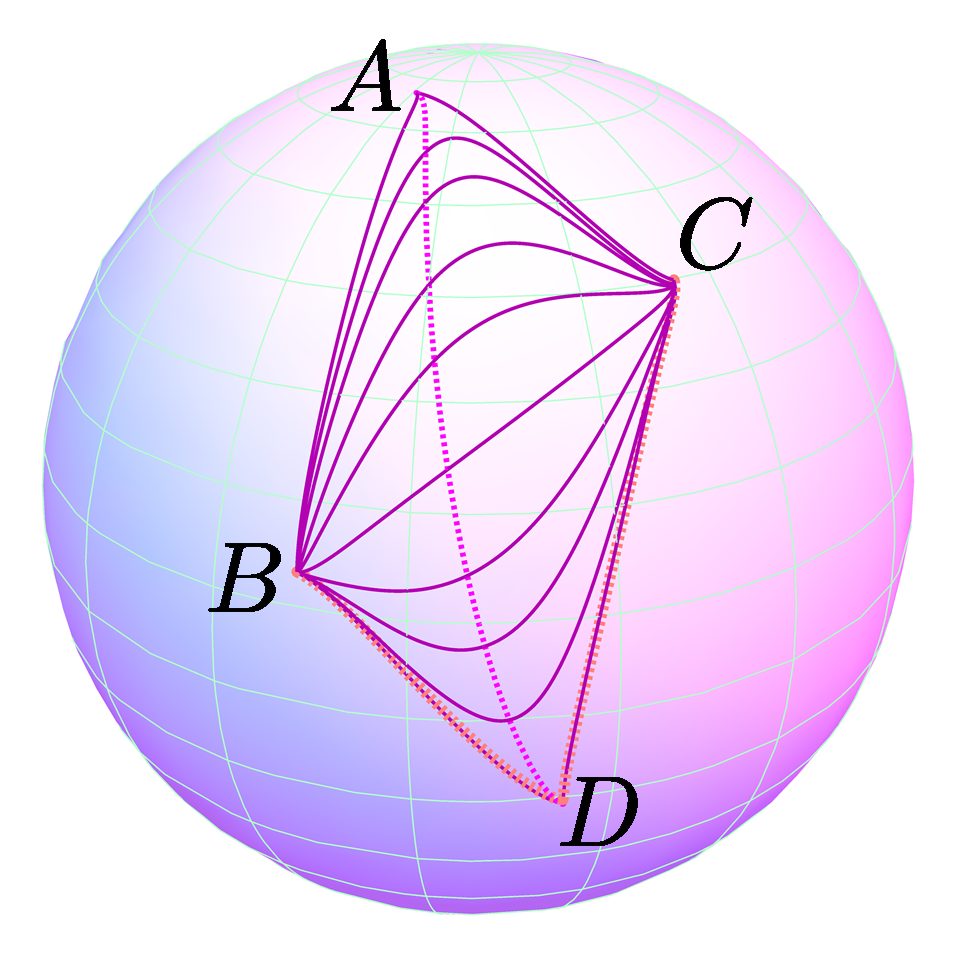}}
\caption{Pictured here is a comparison the results of a collision at impact velocity $u=.995$ in the sphere model with the prediction from parallel transport. On the left we plot the field configuration at various times throughout the collision parametrically in the $\{\phi^i\}$ coordinate plane (by treating the spatial variable as the parameter) with solid purple curves. We identify the field at the origin, $x=0$, throughout the collision with the dot-dashed dark purple line. This is the path taken through field space over the course of the collision by an observer at the center of the collision's rest frame. The solution to the parallel transport problem is shown with dashed lines. Those in pink are lines of constant $\eta$, those in orange are lines of constant $\xi$, i.e. the integral curves of the vector fields $U$ and $W$. The curves that form the boundary of the submanifold are drawn brighter and are overlaid so that they can easily be compared to the results of the collision. 
On the right these results are plotted on the field space manifold embedded in three space (the $\{\phi^i\}$ coordinate lines are shown in light green). The field configuration in the collision problem is again shown in solid purple for a variety of times. The post collision prediction made by parallel transport (that is, the integral curves obtained by completing the parallel transport procedure which yields the remaining two curves that form the boundary of $N$) are shown in dashed orange. The path taken through field space by an observer at the origin is shown in dashed pink. The fact that the boundary of $N$ lines up nearly perfectly with the parametric plots of the initial and final field configuration in the collision problem indicates that there is extraordinarily good agreement between the prediction, computed via parallel transport, and the actual outcome of the collision.}
\label{sphere-results}
\end{figure}

\begin{figure}
{\includegraphics[width=.5\textwidth]{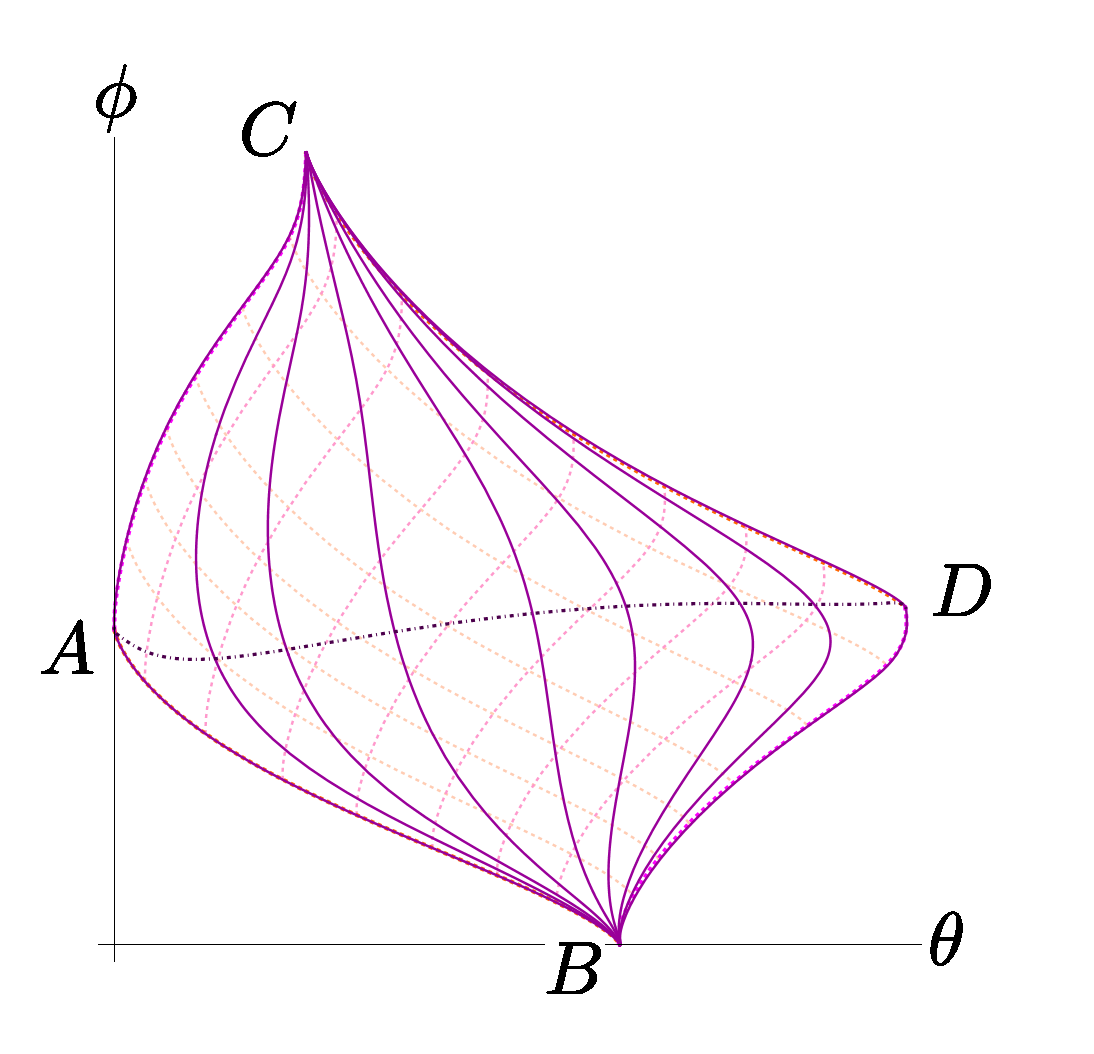}\includegraphics[width=.55\textwidth]{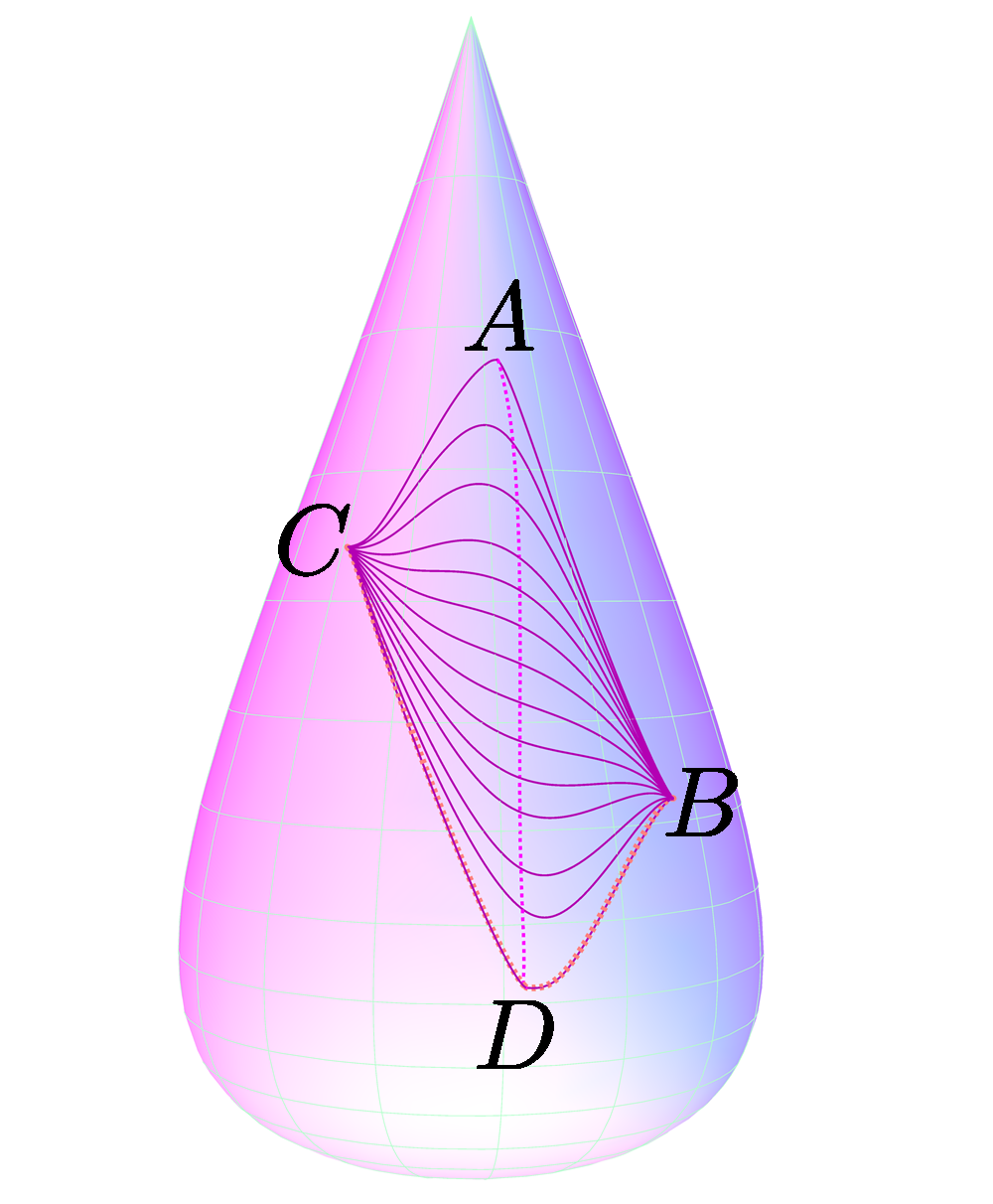}}
\caption{Pictured here is a comparison the results of a collision at impact velocity $u=.995$ in the teardrop model with the prediction from parallel transport. The same coloring scheme is the same as that in Figure \ref{sphere-results}.}
\label{tear-results}
\end{figure}

\begin{figure}
{\includegraphics[width=.5\textwidth]{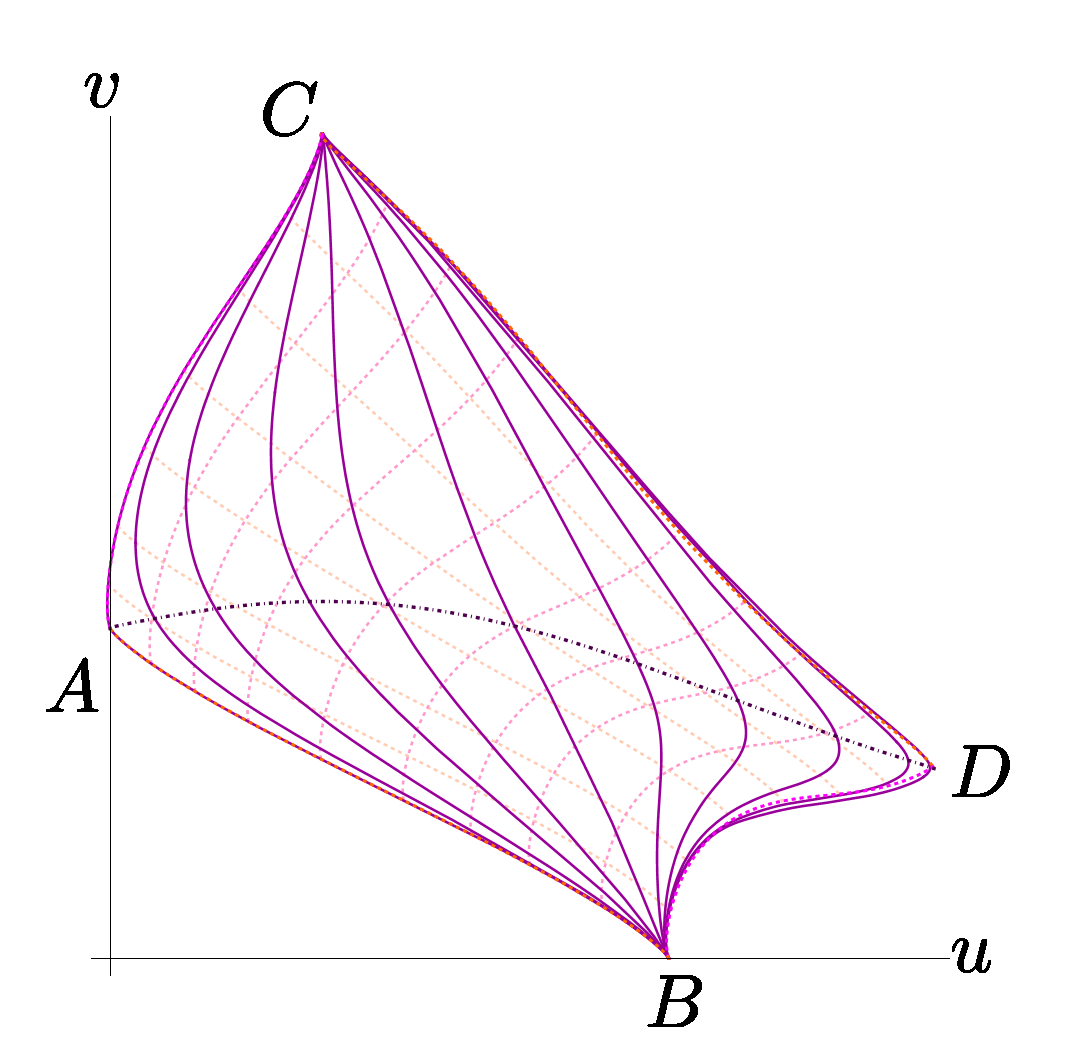}\includegraphics[width=.55\textwidth]{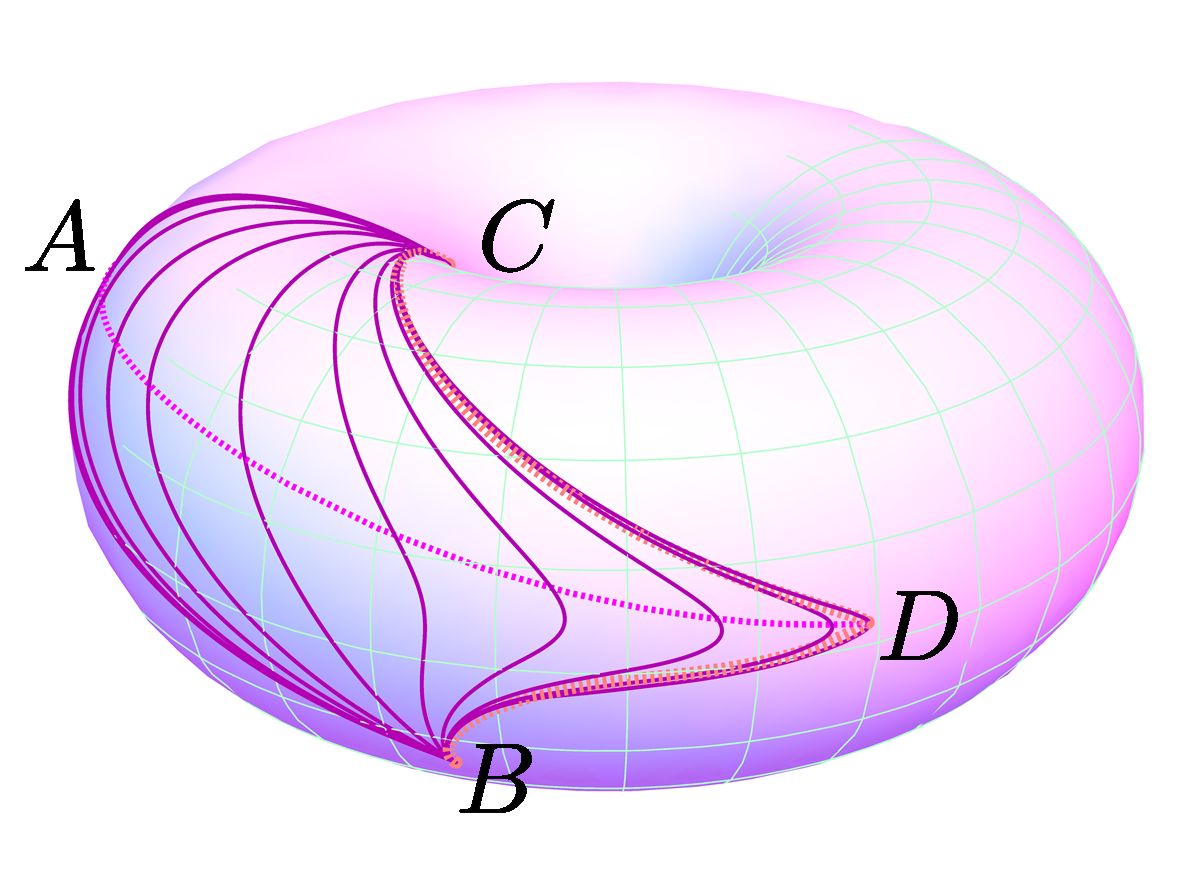}}
\caption{Pictured here is a comparison the results of a collision at impact velocity $u=$ in the teardrop model with the prediction from parallel transport. The same coloring scheme is the same as that in Figure \ref{sphere-results}.}
\label{torus-results}
\end{figure}

The free passage field configuration was indeed fully realized in all three models at sufficiently large impact velocity.  Snapshots of the spatial profile of each field components during such collisions can be found in Figures \ref{snapshots-tear}. Note that each field component's collision region is homogenous, with the precise value predicted by the parallel transport solution, indicated by the contrasting dashed line. Furthermore, the shapes of the outgoing soliton profiles matched the prediction as well. Figures \ref{sphere-results}, \ref{tear-results}, and \ref{torus-results}.

\section{Discussion}

In this paper we pushed the understanding of  bubble universe collisions one step forward by considering the impact of working in the context of a curved field space. Far from an esoteric exercise, this situation arises in prominent contexts such as inflation on the string landscape, in which the relevant fields can be taken to be moduli on Calabi-Yau compactifications. The moduli fields generally span K{\"a}hler manifolds with nontrivial curvature, and so the results of this paper would directly apply.

We have found a simple generalization of the free passage approximation developed in the flat space limit, which admits a satisfying geometrical interpretation. Namely, the free passage evolution is described in field space by a double family of field profiles that interpolate from the two initial solitons along a set of curves, each of whose tangent vectors is parallel transported along the tangent vector field of the other. We have analyzed the conditions under which this curved-field-space free-passage approximation holds and also illustrated its utility in a number of numerical examples.

A natural next step in this program is to include the effects of gravity on bubble collisions, an issue to which we intend to shortly return.

\pagebreak

\section*{Acknowledgements}
We thank Andrew Brainerd, I-Sheng Yang, Eugene Lim and John T. Giblin for helpful discussions. This work was supported by the U.S. Department of Energy under grants DE-SC0011941 and DE-FG02-92ER40699.

\end{document}